\documentclass{article}

\usepackage{arxiv}

\usepackage[utf8]{inputenc} 
\usepackage[T1]{fontenc}    
\usepackage{hyperref}       
\usepackage{url}            
\usepackage{booktabs}       
\usepackage{amsfonts}       
\usepackage{nicefrac}       
\usepackage{microtype}      
\usepackage{amsmath}
\usepackage{amssymb}
\usepackage{cleveref}       
\usepackage{xcolor}
\usepackage{graphicx}
\usepackage[numbers]{natbib}
\usepackage{doi}
\usepackage{diagbox}
\usepackage{rotating}
\usepackage{adjustbox}
\usepackage{algorithm}
\usepackage{algpseudocode}
\usepackage[normalem]{ulem}
\usepackage{subcaption}
\usepackage{color}
\usepackage{ulem}

\usepackage{pgfplotstable}
\usepackage{booktabs}
\pgfplotstableset{
    col sep=comma,
    every head row/.style={before row=\toprule,after row=\midrule},
    every last row/.style={after row=\bottomrule},
    column type/.add={|}{},
}
\newtheorem{definition}{Definition}

\title{Investigation on centrality measures and opinion dynamics in two-layer networks with replica nodes}

\author{ 
  \href{https://orcid.org/0000-0002-1166-7578}{\includegraphics[scale=0.06]{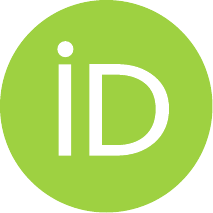}\hspace{1mm}Chi Zhao} \\ 
	Saint Petersburg State University,\\
	7/9 Universitetskaya nab.,\\
	Saint Petersburg, 199034, Russia\\
	\texttt{st081292@student.spbu.ru} \\
	\And
	\href{https://orcid.org/0000-0003-3976-7180}{\includegraphics[scale=0.06]{orcid.pdf}\hspace{1mm}Elena Parilina} \\
	Saint Petersburg State University,\\
	7/9 Universitetskaya nab.,\\
	Saint Petersburg, 199034, Russia\\
	School of Mathematics and Statistics,\\
	Qingdao University,\\
	Qingdao, 266071, PR China\\
	\texttt{e.parilina@spbu.ru} \\
}

\renewcommand{\shorttitle}{Centrality measures and opinion dynamics in two-layer networks with replica nodes}

\hypersetup{
pdftitle={Centrality measures and opinion dynamics in two-layer networks with replica nodes},
pdfauthor={Chi Zhao, Elena Parilina},
pdfkeywords={Opinion dynamics, Centrality measure, General concealed voter model, Zachary's karate club},
}

\begin{document}
\maketitle

\begin{abstract}
  We examine two-layer networks and centrality measures defined on them. We propose two fast and accurate algorithms to approximate the game-theoretic centrality measures and examine connection between centrality measures and characteristics of opinion dynamic processes on such networks. As an example, we consider a Zachary's karate club social network and extend it by adding the second (internal) layer of communication. Internal layer represents the idea that individuals can share their real opinions with their close friends. The structures of the external and internal layers may be different. As characteristics of opinion dynamic processes we mean consensus time and winning rate of a particular opinion. We find significantly strong positive correlation between internal graph density and consensus time, and significantly strong negative correlation between centrality of authoritative nodes and consensus time.
\end{abstract}


\section{Introduction}

The models of opinion dynamics can be divided into macroscopic and microscopic. Macroscopic models including Ising model \cite{mckeehan1925contribution}, Sznajd model \cite{sznajd2000opinion}, voter model \cite{holley1975ergodic}, concealed voter model (CVM) \cite{gastner2018consensus,gastner2019impact}, and macroscopic version of general concealed voter model (GCVM) \cite{zhao2023opinion} examine social networks using sta\-tis\-ti\-cal-physical and probability-theoretical methods to analyze distribution of opinions over time. 

Within GCVM \cite{zhao2023opinion},  it is supposed that the individuals communicate in two layers (internal and external) and can interact in internal or private layer, which makes this model different from CVM, where individuals do not express their true opinions in the internal layer.

Well-known examples of microscopic models of opinion dynamics are DeGroot model \cite{degroot1974reaching}, Friedkin-Johnsen (F-J) model \cite{friedkin1990social}, and bounded confidence models \cite{deffuant2000mixing,rainer2002opinion}. In the F-J model, actors can also factor their initial prejudices into every iteration of opinion \cite{parsegov2016novel}. The models of  opinion dynamics based on DeGroot and F-J models, and Markov chains with possibility to control agents' opinions are proposed in \cite{Bolouki,KAREEVA2023127790,mazalovparilina,sedakovrogov}. The levels of influence and opinion dynamics with the presence of agents with different levels of influence are examined in \cite{Gubanov2,Gubanov1}.

A bounded confidence model (BCM) describes the opinion dynamics, in which agents ignore opinions that are very far from their own opinions \cite{BERNARDO2024111302,noorazar2020recent}. The BCM includes two essential models: the Deffuant-Weisbuch model (D-W) proposed in paper \cite{deffuant2000mixing}, and the Hegselman-Krause (H-K) model introduced in the work \cite{rainer2002opinion}. In the D-W model, two individuals are randomly chosen, and they determine whether to interact according to the bounded confidence \cite{zha2020opinion}.  

The micro version of GCVM is introduced in \cite{zhao2024analysis,10290478}, and the difference between macro and micro versions is the following. In the micro version, we do not need to adjust the simulation program according to different network structures, as long as network structure is given, the program automatically produces simulations. Therefore, we can use this program to simulate  real networks. But for the macro version, we should adjust the corresponding state transition formulae for different network structures. 

Since network structure in GCVM is two-layer, it is interesting to examine how not only this two-layer structure in general, but also network characteristics, e.g. different centrality measures \cite{mazalov2019networking}, affect opinion dynamics and resulting opinion in consensus if it is reached. In this work, we consider two key performance indicators of opinion dynamics, namely, winning rate and consensus time and their relationship with network characteristics.

We focus on centrality measures, considering them as important network characteristics reflecting individual's social power (influence centrality). Centrality measures can be used to identify the most powerful nodes in the network, and information about them is crucial for opinion spreading. The most common centrality measures proposed for one-layer networks are betweenness centrality \cite{freeman1977set}, closeness centrality \cite{bavelas1950communication,FREEMAN1978215,sabidussi1966centrality}, and degree centrality \cite{POWELL201575}. Some centrality measures are based on random walk, such as random walk occupation centrality \cite{sole2016random} which is the frequency of a node in the network being accessed during a random walk; random walk betweenness centrality \cite{newman2005measure}, that is the proportion of the paths through a node to all paths during a random walk. The latter measure does not depend on the shortest path, so it can be considered as a more general one than  betweenness centrality.  Random walk closeness centrality is a variant of closeness centrality \cite{sole2016random}, the computation of which is based on the mean first-passage time (MFPT). The analytical expressions of random-walk based centrality measures can be found in \cite{sole2016random}.

In this paper, we also consider game-theoretic network centralities borrowed from cooperative game theory. A good review on game-theoretic network centrality measures is written by Tarkowski \cite{tarkowski2017game}. The Shapley value \cite{shapley1953value} and the Myerson value \cite{myerson1977graphs} are the cooperative solutions which are proposed as allocations of the total payoff among players based on their marginal contributions. In the paper \cite{suri2008determining}, the authors introduce how to use Shapley value to determine the top-$k$ nodes in the social network. Mazalov et al. in \cite{mazalov2016game} proposed a game-theoretic centrality measure for the weighted graph based on the Myerson value. Mazalov and Khitraya introduced a modified Myerson value for unweighted undirected graphs calculated for a special characteristic function considering not only simple paths but also cycles \cite{mazalov2021modified}. In the next work of these authors \cite{khitraya2024game}, the concept of integral centrality for unweighted directed graph is introduced, and a rigorous mathematical proof that this centrality measure satisfies the Boldi-Vigna axioms \cite{boldi2014axioms} is provided.

In this paper, first we extended our previous work \cite{10.1007/978-3-031-62792-7_21} by considering additional centrality measures and examining the relationship between centrality measures and opinion dynamics in GCVM. Second, we proposed two fast and accurate algorithms to approximate the game-theoretic centrality measures and tested them on randomly generated networks and Zachary's karate club network.
The preliminary conclusion made from this research is that there is a significantly strong positive correlation between internal graph density and consensus time, and significantly strong negative correlation between centrality of ``authoritative'' or most central nodes and consensus time. The proposed algorithms efficiently approximate the Shapley and Myerson values in randomly generated networks with high accuracy. Additionally, our algorithms successfully determine the two most influential nodes in the Zachary's karate club network.

The rest of this paper is organized as follows. Section 2 introduces multi-layer network with replica nodes. Network properties and particular centrality measures are discussed in Section 3, where we also introduce algorithms to approximate the Shapley and Myerson values. Section 4 describes simulation experiments and results of these experiments. We briefly conclude and discuss our future work in Section 5.

\section{Multi-layer networks with replica nodes}

\subsection{Multi-layer network}

A multilayer network is a network formed by several networks that evolve and interact with each other \cite{bianconi2018multilayer}.

\subsection{Replica nodes}

In a multilayer network with replica nodes there is a one-to-one mapping of the nodes in different layers and corresponding nodes are called replica nodes. Since there is a one-to-one mapping between the nodes in different layers, every layer is formed by the same number of nodes  \cite{bianconi2018multilayer}.

\subsection{Two-layer networks with replica nodes}

We use the following notations to define a two-layer (external and internal layers) network with replica nodes:
\begin{itemize}
\item $N$: number of nodes in each layer;
\item $a_i=(a_i^E,a_i^I)$: one-to-one mapping of node $i$ in the external and internal layer, where $a_i^E$ ($a_i^I$) is a representation of node $i$ in the external (internal) layer;
\item $G_E(\mathcal{V}_E, \mathcal{E}_E)$: predefined external network, where $\mathcal{V}_E=\{a_i^E\}$ and $\mathcal{E}_E$ represent a set of individuals and set of edges in the external layer, respectively; 
\item $G_I(\mathcal{V}_I, \mathcal{E}_I)$: predefined internal network, where $\mathcal{V}_I=\{a_i^I\}$ and $\mathcal{E}_I$ represent a set of individuals and set of edges in the internal layer, respectively;
\item $\mathcal{E}_C=\{(a_i^E,a_i^I)|i=1,\ldots,N\}$: edges connecting replica nodes.
\end{itemize}

A two-layer network with $N$ individuals/agents can be defined as\footnote{Opinion dynamic models in two-layer networks with replica nodes was introduced in \cite{10290478} and further discussed in \cite{zhao2024analysis} and \cite{10.1007/978-3-031-62792-7_21}.}:
  \begin{equation}
  \label{eq:two-layer-network-def}
    G(\mathcal{V},\mathcal{E}),
  \end{equation}
where $\mathcal{V}=\mathcal{V}_E\cup \mathcal{V}_I$, $|\mathcal{V}_E|=|\mathcal{V}_I|=N$, and $\mathcal{E}=\mathcal{E}_E\cup \mathcal{E}_I\cup \mathcal{E}_C$.

\subsection{Two-layer network simplification}
\label{sec:simplification}
	
Two-layer network $G(\mathcal{V},\mathcal{E})$ is composed of external network $G_E(\mathcal{V}_E, \mathcal{E}_E)$, internal network $G_I(\mathcal{V}_I, \mathcal{E}_I)$ with the set of edges $\mathcal{E}_C$ connecting nodes between layers. Two-layer networks can also be represented by an adjacency matrix. The adjacency matrix of a two-layer network is a block matrix, where the diagonal blocks are the adjacency matrices of the external and internal layers, and the off-diagonal blocks are the adjacency matrices of the connections between external and internal layers.
	
The adjacency matrix of $G(\mathcal{V},\mathcal{E})$ is as follows:	
	\begin{equation}
		\label{eq:adjacency-matrix}
		A = \begin{bmatrix}
			A_{EE} & A_{EI} \\
			A_{IE} & A_{II}
		\end{bmatrix}
	\end{equation}
where $A_{EE}$ is the adjacency matrix of the external layer, $A_{EI}$ is the adjacency matrix of the connections between external and internal layers, $A_{IE}$ is the adjacency matrix of the connections between  internal and external layers, and $A_{II}$ is the adjacency matrix of the internal layer. For undirected graphs in both layers, the adjacency matrix $A$ is symmetric.

Keeping in mind an binary opinion dynamics model presented in \cite{zhao2023opinion}, we define the rates of coping opinions from one node to another:
\begin{itemize}
 	\item $\pi_{c_e}$: external copy rate with which an individual is copying opinion of his/her external neighbor if they both are randomly chosen;
 	\item $\pi_{c_i}$: internal copy rate with which an individual is  copying opinion of his/her internal neighbor;
 	\item $\pi_e$: externalization rate with which an individual is behaving as a hypocrisy\footnote{By hypocrisy we mean a node having different opinions in external and internal layers.} choosing to publicly express his/her internal opinion;
 	\item $\pi_i$: internalization rate with which an individual being hypocrisy accepts his/her external opinion.
 \end{itemize}
	
We propose a way how to transform a two-layer network $G(\mathcal{V},\mathcal{E})$ with the given parameters $\pi_{c_e},\pi_{c_i},\pi_e,\pi_i$ of opinion dynamics to a one-layer weighted network. We can define a matrix of weights as follows:
	\begin{equation}
		\label{eq:fusion-rule}
		W'= \pi_{c_e} \cdot A_{EE} + \pi_{c_i} \cdot A_{II} + \pi_{i} \cdot \Lambda_{E} + \pi_{e} \cdot \Lambda_{I},
	\end{equation}
where $\Lambda_E$ and $\Lambda_I$ are $N\times N$ diagonal matrices, and the elements on the diagonal represent the degrees of nodes in external and internal layers, respectively.
Furthermore,  we use $w'_{ij}$ to represent the elements of matrix $W'$ and define a new weighted network $G'(\mathcal{V}',\mathcal{E}',W')$, where $\mathcal{V}'=\{1,2,\ldots,N\}$ is the set of nodes,  $\mathcal{E}' = \{(i,j) \mid w'_{ij} \neq 0, i,j \in \mathcal{V}'\}$ is the set of edges, and $W'$ is the matrix of weights.
	
Based on a new weighted network $G'(\mathcal{V}',\mathcal{E}',W')$, we proposed two game-theoretic centrality measures which will be discussed in  Section~\ref{sec:game-centrality}.

\subsection{Zachary's karate club in a two-layer network setting}

\begin{figure}[!htbp]
	\centering
		\includegraphics[scale=0.5, clip, trim=0cm 1.5cm 0cm 7.5cm]{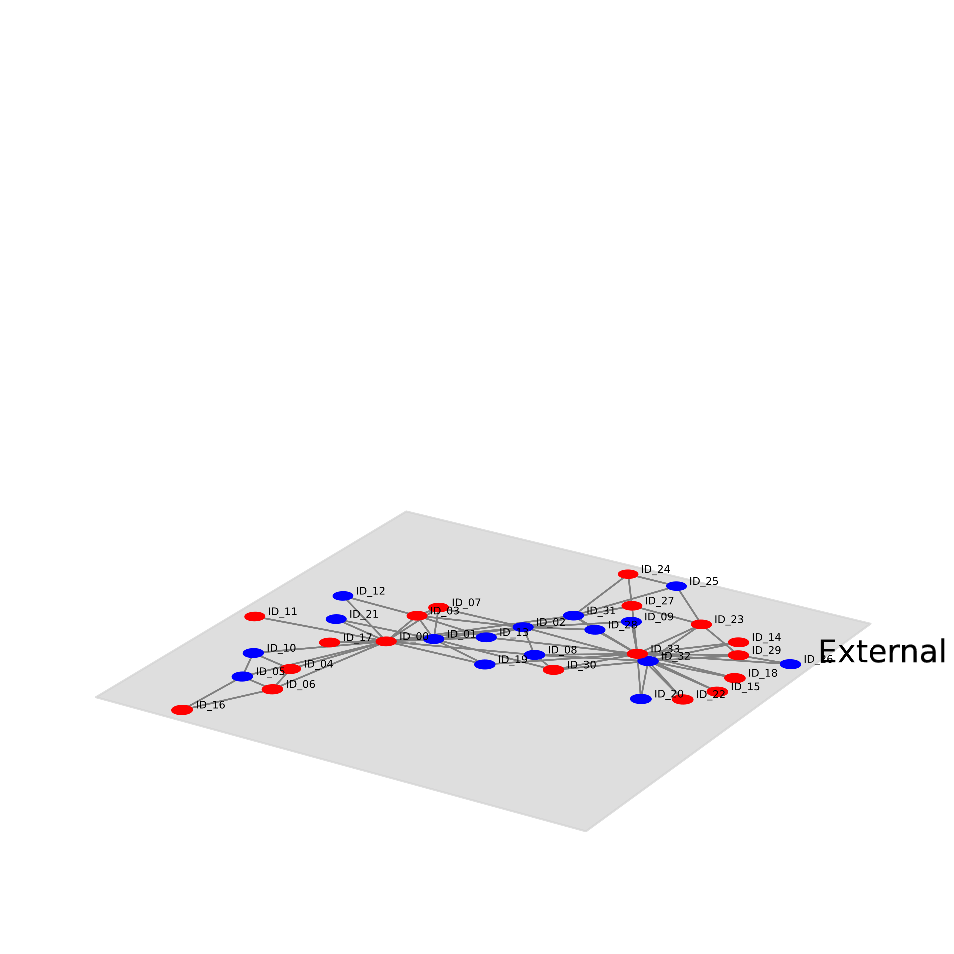}
		\caption{One-layer Zachary's karate club network}
		\label{fig:karate-bvm}
\end{figure}

As an example of a social network, we consider Zachary's karate club network representing friendship relations among 34 members of a karate club at the US university in the 1970s \cite{zachary1977information}. The study became famous in data and network analytical literature since it highlighted a conflict between manager (Node 0) and director (Node 33), which eventually led to the split of the club into two groups. One-layer Zachary's karate club network is represented in Fig.~\ref{fig:karate-bvm}. The blue and red colors of nodes represent two opinions in the social network.

\begin{figure}[!htbp]
  \begin{subfigure}[normla]{.45\linewidth}
    \includegraphics[scale=0.33]{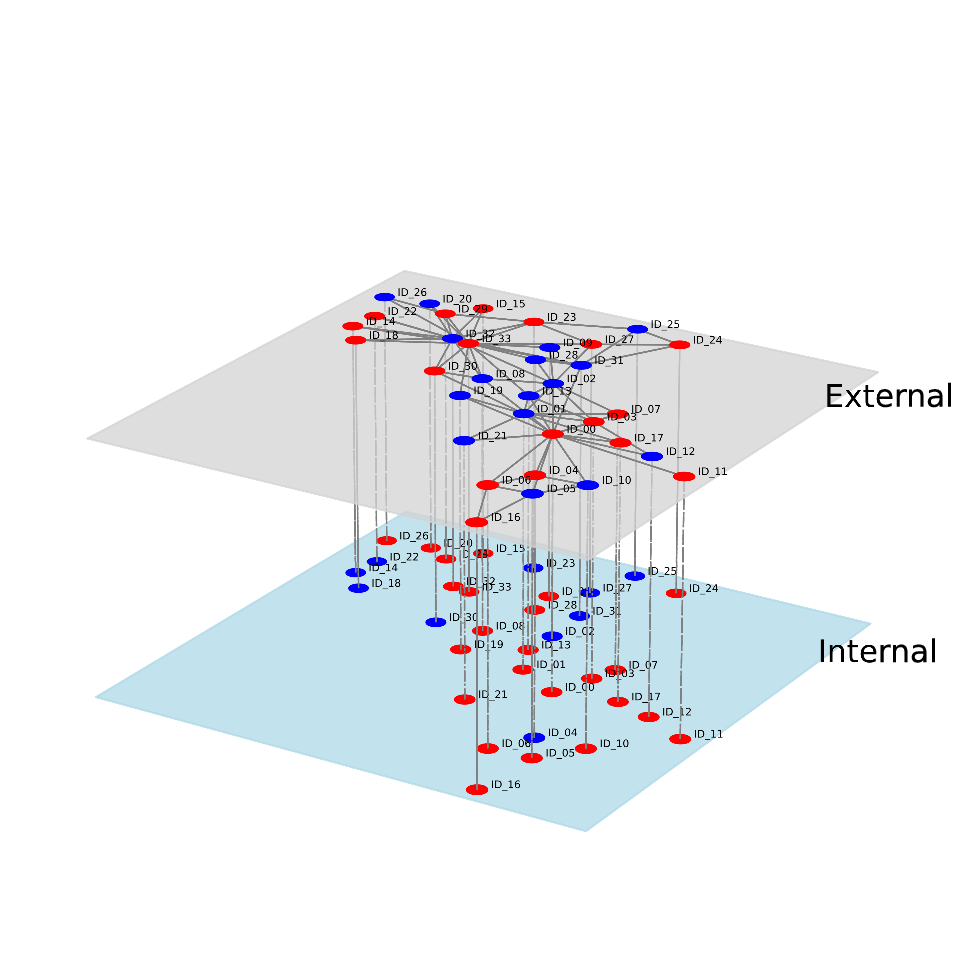}
    \subcaption{CVM}
    \label{fig:karate-cvm}
  \end{subfigure}
  \begin{subfigure}[normla]{.45\linewidth}
    \includegraphics[scale=0.33]{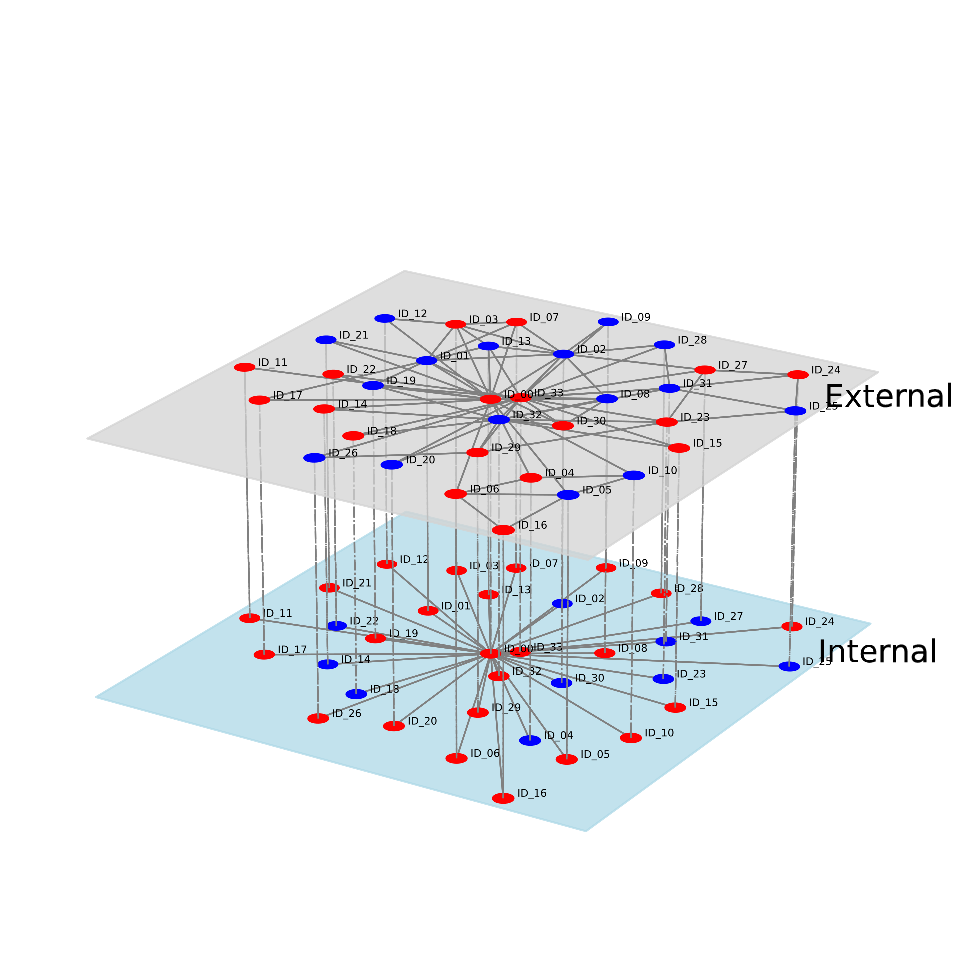}
    \subcaption{GCVM}
    \label{fig:karae-gcvm}
  \end{subfigure}
  \caption{Two-layer networks used in CVM and GCVM: (a) CVM: two-layer network with external Zachary's karate club and empty internal layer, (b) GCVM: two-layer network with external Zachary's karate club and star internal layer.}
  \label{fig:different-models}
\end{figure}

Fig.~\ref{fig:different-models} shows how one-layer Zachary's Karate Club network can be extended into a two-layer network if we add an internal layer of communication between agents. If we consider binary opinion dynamics models, in the concealed voter model (CVM) \cite{gastner2018consensus,gastner2019impact},  the nodes in the internal layer are not connected, i.e. internal layer is represented by an empty graph (see Fig.~\ref{fig:karate-cvm}), while in the general concealed voter model (GCVM) \cite{zhao2024analysis,10290478,zhao2023opinion}  there may be nonempty network representing internal communication of agents. In Fig.~\ref{fig:karae-gcvm} we represent a star internal structure. The colors in Fig~\ref{fig:different-models} represent individuals'\ opinions. 

The color, blue or red, is randomly initialized for the given parameters, these are (i) probability of having red initial opinion for the basic voter model on  one-layer network, (ii) probabilities of having red initial opinion in external, internal, and in both layers for CVM and GCVM models. 

In the basic voter model (BVM, see \cite{holley1975ergodic}), there is only one layer, as shown in Fig.~\ref{fig:karate-bvm}, and each individual holds one of two opinions (red or blue). At each step, a random individual selects a random neighbor and adopts his/her neighbor's opinion with the copying rate $\pi_{c_e}$. This process repeats until everyone in the network holds the same opinion, i.e. reaching a consensus. The opinion dynamics process in CVM \cite{gastner2018consensus,gastner2019impact} and GCVM \cite{zhao2024analysis,10290478,zhao2023opinion,10.1007/978-3-031-62792-7_21} is implemented on a two-layer network (see  Fig.~\ref{fig:karate-cvm} and \ref{fig:karae-gcvm}). In the CVM, individuals in the internal layer do not communicate, while in GCVM, individuals in the internal layer can communicate. The opinion dynamics process in CVM and GCVM is similar to BVM, but have more options: in the CVM, individuals can his/her internal opinion publicly with externalization rate $\pi_{e}$ or accept his/her external opinion with internalization rate $\pi_{i}$. These two options do not exist in the BVM, while in GCVM, individuals can copy their internal neighbors' opinion with internal copy rate $\pi_{c_i}$, which is not allowed in the CVM.

%
%
%

\section{Centrality measures in one- and two-layer networks}

In this section, we represent several centrality measures. Some of them are defined for one-layer networks and can be applied for two layers separately, some of them take into account the multi-layer structure of a network. We also introduce game-theoretical centrality measures and provide an algorithm to calculate their approximation when the network contains large number of nodes. 

\begin{definition} 
A pairwise average shortest path for an external layer $d_E$ in a two-layer network is
 \begin{equation}
  	\label{eq:avg_shortest_path_length}
  	d_E=\sum_{s,t\in \mathcal{V}_E} \frac{d_E(s,t)}{n_E(n_E-1)},
  \end{equation}
where $d_E(s,t)$ is a length of the shortest path between nodes $s$ and $t$ in the external layer, $\mathcal{V}_E$ is a set of nodes in the external layer, $n=|\mathcal{V}_E|$ is a number of nodes in the external layer. Similarly, we can define pairwise average shortest path for an internal layer denoted by $d_I$.
\end{definition}

\begin{definition}
A graph density is a ratio of the number of edges $|\mathcal{E}|$ with respect to the maximal number of edges. Since internal layer is represented by an undirected graph, we define an internal graph density as in \cite{Densegra41:online}:
\begin{equation}
    D_I=\frac{2|\mathcal{E}_I|}{|\mathcal{V}_I|(|\mathcal{V}_I|-1)}.
\end{equation}
\end{definition}

\subsection{Classical centrality measures}

In this section we briefly introduce some (most well-known) centrality measures defined for one-layer networks. In the rest of the paper, we use $V$ to denote the set of nodes in a one-layer network.

\subsubsection{Betweenness centrality}

Betweenness centrality of a node introduced in \cite{freeman1977set}  gives the number of geodesics between all nodes that contain this node. It reflects the level of node participation in the dissemination of information between other nodes in a graph. It is calculated by the formula:
\begin{equation}
    \label{eq:betweenness}
    C_b(v) = \frac{1}{n_b}\sum\limits_{s,t\in V}\frac{\sigma_{s,t}(v)}{\sigma_{s,t}},
\end{equation}
where $\sigma_{s,t}$ indicates the number of shortest paths between nodes $s$ and $t$, and $\sigma_{s,t}(v)$ is the number of shortest paths between nodes $s$ and $t$ containing node $v$. Normalization coefficient is $n_b=(|V|-1)(|V|-2)$ for $v\notin \{s,t\}$, otherwise $n_b=|V|(|V|-1)$, where $|V|$ is the number of nodes in a one-layer network \cite{mazalov2019networking}. If $s=t, \sigma_{s,t}=1$ and if $v\in \{s,t\}$, then $\sigma_{s,t}(v)=0$.

\subsubsection{Group betweenness centrality}

Group betweenness centrality measure indicates a proportion of shortest paths connecting pairs of nongroup members that pass through the group (see \cite{everett1999centrality}), and it is defined by formula:
\begin{equation}
    \label{eq:group-betweenness}
    C_{gb}(X) = \frac{1}{n_{gb}}\sum\limits_{s,t\in V\setminus X }\frac{\sigma_{s,t}(X)}{\sigma_{s,t}},
\end{equation}
where $\sigma_{s,t}(X)$ is the number of shortest paths between nodes $s$ and $t$ passing through some nodes in group $X$. Normalization coefficient is $n_{gb}=(|V|-|X|)(|V|-|X|-1)$, where $|X|$ is the number of nodes in  group $X$.

\subsubsection{Closeness centrality}

In a connected graph, closeness centrality of node $u$ is the reciprocal of a sum of lengths of the shortest paths between $u$ and all other nodes in the graph \cite{bavelas1950communication,FREEMAN1978215,sabidussi1966centrality}. When calculating closeness centrality, its normalized form is usually referred to as the one representing the average length of the shortest path instead of their sum, and it is calculated like this: 
\begin{equation}
    \label{eq:closeness}
    C_c(u) = \frac{n_c}{\sum_{v\in V\setminus \{u\}}d(v,u)},
\end{equation}
where normalization coefficient is $n_c=|V|-1$.

\subsubsection{Group closeness centrality}

Group closeness centrality is the reciprocal of the sum of the shortest distances from the group to all nodes outside the group \cite{everett1999centrality,everett2005extending,zhao2014measuring}. It is calculated as follows:
\begin{equation}
    \label{eq:group-closeness}
    C_{gc}(X) = \frac{n_{gc}}{\sum_{v\in V\setminus X}d(v,X)},
\end{equation}
where $d(v,X)$ is the shortest distance between group $X$ and $v$. Normalization coefficient is $n_{gc}=|V-X|$.

\subsubsection{Degree centrality}

Degree centrality of node $v$ \cite{POWELL201575} is defined as
\begin{equation}
    \label{eq:degree-centrality}
    C_d(v) = \frac{v_d}{n_d},
\end{equation}
where $v_d$ is a degree of node $v$, and normalization coefficient is $n_d=|V|-1$.

\subsubsection{Group degree centrality}

Group degree centrality is the number of nodes outside the group  connected with the nodes from this group \cite{everett1999centrality,everett2005extending}. Normalized group degree centrality for group $X$ is given by the formula:
\begin{equation}
    \label{eq:group-degree-centrality}
    C_{gd}(X) = \frac{|\{v_i\in V \setminus X|v_i \text{ is connected to } v_j \in X\}|}{n_{gd}},
\end{equation}
where normalization coefficient is $n_{gd}=|V|-|X|$.

\subsection{Random walk based centralities}

The second group of centrality measures is based on random walks, 
 simple dynamical process that can occur on a network. Random walks can be also used to approximate other types of diffusion processes \cite{chung1997spectral,newman2010networks,sole2016random}. 
 
\subsubsection{Random walk occupation centrality}

The random walk occupation centrality \cite{sole2016random} of node $v$ is the probability of that node $v$ being visited by a random walker during an infinitely long walk, and it is defined as
\begin{equation}
    \label{eq:random-walk-occupation}
    C_{rwoc}(v) = \lim_{t\to \infty}\frac{n_v(t)}{t},
\end{equation}
where $n_v(t)$ is the number of times node $v$ is visited by a random walker during time interval $t$. Different exploration strategies can be used to calculate the occupation centrality, we use the uniform exploration strategy in this paper (i.e. each node jumps to its neighbor with the equal probability). In the weighted networks, jumping probabilities are proportional to the weights of the edges.

The analytical expressions of random walk occupation centrality with a uniform exploration strategy in interconnected multilayer networks are presented in \cite{sole2016random}.

\subsubsection{Random walk betweenness centrality}

The most common betweenness is the shortest path betweenness \cite{freeman1977set,sole2016random}, where the centrality of a node $v$ is relative to the number of shortest paths between all pairs of nodes passing through $v$. However, in real networks, entities (rumors, messages, or internet packets) that travel the network do not always follow the shortest path \cite{freeman1991centrality,sole2016random,stephenson1989rethinking}. Therefore, the random walk betweenness centrality of node $v$ is defined as the number of random walks between any pair $(s,d)$ of nodes that pass through node $v$ \cite{newman2005measure}:
\begin{equation}
  C_{rwbc}(v) = \frac{1}{n_{rwbc}} \sum_{\substack{s, t \in V \\ s \neq t \\ v \neq s, v \neq t}} \mathbf{1}_{v \in \text{Path}_{s \to t}},
\end{equation}
where $n_{rwbc}=2N (N-1)$ is the normalization coefficient. The indicator function $\mathbf{1}_{v \in \text{Path}_{s \to t}}$ is equal to 1 if node $v$ is in the path between nodes $s$ and $t$, and 0 otherwise. The $\text{Path}_{s \to t}$ is the random path between nodes $s$ and $t$ in the network. Repeating the random walk process few times to get different random paths we get the average random walk betweenness centrality.

It will be useful to get the analytical expression of random walk betweenness centrality for nodes by absorbing random walk, where the absorbing state is selected to be the destination node $d$ \cite{newman2010networks,newman2005measure}. An extended analytical expression of random walk betweenness centrality for interconnected multilayer networks can be found in \cite{sole2016random}.

\subsubsection{Random walk closeness centrality}
A variant of closeness centrality is random walk closeness centrality, the computation of which is based on the mean first-passage time (MFPT). The MFPT is defined as the average number of steps to reach node $d$ starting from a given node $s$. The lower average MFPT indicates that a node is on average more quickly accessible from other nodes. Therefore, a node with a lower average MFPT to all other nodes is considered more ``central'' in the network. The random walk closeness centrality is defined as the reciprocal of the average MFPT, and it is calculated by the formula:
\begin{equation}
    \label{eq:random-walk-closeness}
    C_{rwcc}(v) = \frac{n-1}{\sum_{u\in V\setminus \{v\}}\tau_{uv}},
\end{equation}
where $\tau_{uv}$ is the MFPT from node $u$ to node $v$. The MFPT matrix can be calculated analytically by means of Kemeny-Snell fundamental matrix $Z$ \cite{lovasz1993random,zhang2011mean} or by means of absorbing random walks \cite{kemeny1969finite,newman2010networks}. 

The analytical expressions of random walk closeness centrality in interconnected multilayer networks can be found in \cite{sole2016random}.

\subsection{Game-theoretic centrality measures}\label{sec:game-centrality}

\subsubsection{Shapley value based centrality}

The Shapley value is a solution concept in cooperative game theory introduced by Lloyd Shapley in 1953 \cite{shapley1953value}. It is a measure of the average marginal contribution of a player to all possible coalitions. Shapley value is a solution concept assigning a singleton solution to the players as allocation given by formula:
\begin{equation}
    \label{eq:shapley-value}
    \phi(i) = \sum_{S\subseteq N\setminus \{i\}}\frac{|S|!(n-|S|-1)!}{n!}(v(S\cup \{i\})-v(S)),
\end{equation}
where $S \subseteq V$ represents a coalition, the value of coalition $S$ can be denoted by $ v(S) $. We define characteristic function $v(S)$ as half of the sum of the weighted degrees of all nodes in the subgraph induced by $S$, that is, 
\begin{equation}
  \label{eq:shapley-coalition-value}
  v(S) = \frac{1}{2} \sum_{\{i, j\} \subseteq S} W(i, j), 
\end{equation}
where $ W(i, j)$ is the weight of the edge between nodes $ i $ and $ j $ within the subgraph induced by $ S $, coefficient $\frac{1}{2}$ is used for correction in case of an undirected graph, each edge is counted twice when summing up over all pairs of nodes.

\begin{algorithm}[!htbp]
	\caption{Calculation of the Shapley Values based on weighted graph}
	\label{alg:shapley-value}
	\begin{algorithmic}[1] 
		\Require A graph $G(V,E,W)$ with $n = |V|$ nodes
		\Ensure Shapley value component $\phi(i)$ for each node $i \in V$
		
		\ForAll{nodes $i \in V$}
		\State Initialize $\phi(i) \leftarrow 0$
		\EndFor
		
		\ForAll{nodes $i \in V$}
		\ForAll{subsets $S \subseteq V \setminus \{i\}$}
		\State Compute $v(S) \leftarrow \sum_{\{j,k\} \subseteq S} W(j,k)$ within subgraph induced by $S$
		\State Compute $v(S \cup \{i\})$ within subgraph induced by $S \cup \{i\}$
		\State $\Delta v(S, i) \leftarrow v(S \cup \{i\}) - v(S)$
		\State coeff $\leftarrow \frac{|S|! \cdot (n - |S| - 1)!}{n!}$
		\State $\phi(i) \leftarrow \phi(i) + \text{coeff} \cdot \Delta v(S, i)$
		\EndFor
		\EndFor
		
		\Return $\phi(i)$ for all $i \in V$
	\end{algorithmic}
\end{algorithm}

Algorithm~\ref{alg:shapley-value} describes how to calculate the Shapley value based on a weighted graph. However, the Shapley value is computationally expensive,  especially, for large networks with the large number of coalitions.\footnote{For a network with $n$ nodes, the total number of coalitions is equal to $2^n$.} Therefore, we propose a new approach to calculate an approximated Shapley value described in the next section.

\subsubsection{Approximated Shapley value}

According to the fact that the influence from other nodes is decreasing with an increase of the path length, we propose several ideas to fasten the calculation of the Shapley value:
\begin{enumerate}
  \item[1.] \textit{Depth limitation}: By limiting the depth of the  reachable nodes considered, the number of subsets that need to be considered is reduced. 
  \item[2.] \textit{Local subset iteration}: Iterating over subsets only within the reachable nodes, rather than the entire graph, decreases the number of iterations.
  \item[3.] \textit{Neighbor size sampling}: For a large number of neighbors, computational complexity can be reduced by random sampling, thereby decreasing the number of subsets iterated over.
\end{enumerate}

We first define $\psi(i,d_{max})$ as the set of reachable nodes of node $i$ up to given depth $d_{max}$ excluding node $i$. We can calculate an approximated Shapley value of node $i$ based on set $\psi(i,d_{max})$ by formula:
\begin{equation}
	\label{eq:shapley-value-approximation}
	\phi_a(i) = 
	\begin{cases}
		\sum_{S\subseteq \psi(i,d_{max})}\frac{v(S\cup \{i\})-v(S)}{2^{|\psi(i,d_{max})|}}  &\text{ if } |\psi(i,d_{max})| < m, \\[10pt]
		\beta\sum_{S\subseteq \psi(i,d_{max})}\frac{v(S\cup \{i\})-v(S)}{2^{|\psi(i,d_{max})|}}  &\text{ if } |\psi(i,d_{max})| \geq m,
	\end{cases}
\end{equation}
where $\beta=\frac{|\psi(i,d_{max})|+1}{m+1}$ is a scaling factor, and $m$ is the maximal number of reachable nodes considered.

For $|\psi(i,d_{max})|\geq m$, we create a random sample of $m$  nodes from set $\psi(i,d_{max})$ several times and calculate the components of the Shapley value based on these samples. The sampling time $H_{|\psi(i,d_{max})|,m}$ is given by formula~\eqref{eq:sampling-times} (see \cite{PlyaEineWI}):
\begin{equation}
	\label{eq:sampling-times}
	H_{|\psi(i,d_{max})|,m} = \left(\frac{|\psi(i,d_{max})|+\frac{1}{2}}{m} - \frac{1}{2}\right)\left(\ln |\psi(i,d_{max})|+\gamma\right)+\frac{1}{2},
\end{equation}
where $\gamma \approx 0.5772156649$ is the Euler-Mascheroni constant. This formula is the mathematical expectation of the number of samples for collecting $m$ nodes from set $\psi(i,d_{max})$ until all reachable nodes are collected.\footnote{We can consider this problem as the generalized coupon collector's problem \cite{PlyaEineWI}.}

Equation~\eqref{eq:shapley-value-approximation} gives a good approximation of the Shapley value, especially when the density of the graph is not very high (less than 0.7). We should highlight that our algorithm provide an accurate estimation of the ratio of the approximated component of the Shapley value to the sum of all its components. Knowing this ratio and exact value $v(N)$, it is easy to calculate the approximated Shapley value using the scaling factor $\xi$:
\begin{equation}
	\label{eq:shapley-value-approximation-factor}
	\xi = \frac{v(N)}{\sum_{i\in V}\phi_a(i)}.
\end{equation}

The steps of calculation of the approximated Shapley value in weighted graphs is described in Algorithm~\ref{alg:shapley-approximation}. We use the same characteristic function as in the original Shapley value calculation, but we reduce the number of calculations following the ideas listed above.

We can get a more accurate approximated Shapley value by multiplying its components $\phi_a(i)$ given by equation \eqref{eq:shapley-value-approximation} with  factor $\xi$ defined by \eqref{eq:shapley-value-approximation-factor}.
We include this step in the benchmark of Algorithm \ref{alg:shapley-approximation}. The results of its work are presented in  Section~\ref{sec:centrality-experiments-description}.

\begin{algorithm}[!htbp]
\caption{Calculation of approximated Shapley Value in weighted graph}
\label{alg:shapley-approximation}
\begin{algorithmic}[1]
  \Require A weighted graph $G = (V, E, W)$, depth limit $d_{max}$, maximal size $m$ of the set of reachable considered
  \Ensure Approximated Shapley value $(\phi_a(i), i \in V)$
  \State Initialize $\phi_a(i) \gets 0$ for each $i \in V$
  \For{$i \in V$}
  \State $\psi(i,d_{max}) \gets$ Calculate or retrieve all $reachable~nodes$ of $i$ up to depth $d_{max}$
      \If{$|\psi(i,d_{max})| < m$}
          \For{each subset $S \subseteq \psi(i,d_{max}) \setminus \{i\}$}
              \State Compute $v(S) \leftarrow \sum_{\{j,k\} \subseteq S} W(j,k)$ within subgraph induced by $S$
              \State Compute $v(S \cup \{i\})$ within subgraph induced by $S \cup \{i\}$
              \State $\Delta v(S, i) \leftarrow v(S \cup \{i\}) - v(S)$
              \State $\phi_a(i) \gets \phi_a(i) + \Delta v(S, i)$
          \EndFor
          \State coeff $\leftarrow \frac{1}{2^{|\psi(i,d_{max})|}}$
          \State $\phi_a(i) \gets \phi_a(i) \cdot \text{coeff}$, normalize $\phi_a(i)$ based on the number of subsets
      \Else
      \State Pick up $m$ nodes randomly from $\psi(i,d_{max})$ and repeat $H_{|\psi(i,d_{max})|,m}$ times
          \For{ i = 1 to $H_{|\psi(i,d_{max})|,m}$}
              \State $s_{reachable} \gets $ Randomly select a sample of $m$ nodes from $\psi(i,d_{max})$, 
              \For{each subset $S \subseteq s_{reachable} \setminus \{i\}$}
                  \State Calculate $v(S)$ and $v(S \cup \{i\})$ as before
                  \State $\Delta v(S, i) \leftarrow v(S \cup \{i\}) - v(S)$
                  \State $\phi_a(i) \gets \phi_a(i) + \Delta v(S, i)$
              \EndFor
          \EndFor
            \State coeff $\leftarrow {1}/{2^{|\psi(i,d_{max})|}}/H_{|\psi(i,d_{max})|,m} \cdot \frac{|\psi(i,d_{max})|+1}{m+1}$
            \State $\phi_a(i) \gets \phi_a(i) \cdot \text{coeff}$
      \EndIf
  \EndFor
  \State Define  scaling factor $\xi \gets \frac{v(N)}{\sum_{i\in V}\phi_a(i)}$ \Comment{For accurate results}
  \State \Return $\phi_a(i) \gets \xi \cdot \phi_a(i)$ for all $i \in V$
\end{algorithmic}
\end{algorithm}

\subsubsection{Myerson value based centrality} The Myerson value was introduced by Roger Myerson in 1977 \cite{myerson1977graphs}, and it is an allocation rule  when players are connected by a network structure. By modifying the calculation method of the Shapley value, Myerson takes into account connections in the network, thereby reflecting the influence of network structure on the cooperative game. Consider a game where graph $G$ is a tree, which consists on $N$ nodes and characteristic function is determined by the scheme proposed in \cite{jackson2008social}. Every direct connection gives to coalition $S$ a value $r$, where $0\leq r\leq 1$. Players also obtain an impact from non-direct connections. This kind of impact will decrease with the increase of the path length. The characteristic function is defined as follows \cite{mazalov2016game}:
\begin{equation}
  \label{eq:myerson-jackson}
    v(S) = a_1r + a_2r^2 + \dots + a_kr^k + \dots + a_Lr^L = \sum_{k=1}^{L} a_k r^k,
\end{equation}
where $L$ is a maximal distance between two nodes in the coalition; $a_k$ is the number of paths of length $k$ in this coalition; $v(i) = 0, \forall i \in N.$

Mazalov et al. in \cite{mazalov2014generating} proved that an allocation rule, that is, the Myerson value for unweighted graphs:
\begin{equation}
	\label{eq:myerson-mazalov}
	Y_i (v,g) = \frac{\sigma_1(i)}{2} r + \frac{\sigma_2(i)}{3} r^2 + \cdots + \frac{\sigma_L(i)}{L+1} r^L = \sum_{k=1}^{L} \frac{\sigma_k(i)}{k+1} r^k,
\end{equation}
where $\sigma_k(i)$ is a number of the paths of the length $k$ including node $i$. The same approach to define an allocation can be applied to weighted graphs by converting the weight of the edge to the number of paths between two nodes, i.e. by converting a weighted graph to a multigraph \cite{mazalov2016game}.

\subsubsection{Approximated Myerson value} 

However, calculation of the Myerson value is also computationally expensive, especially, for large networks. We can consider the rule of ``six degrees separation'' \cite{milgram1967small}, implementing the idea that any two persons in the world who do not know each other only need a few intermediaries to establish a contact. Based on this idea, we can reduce the computational expense by limiting the maximal depth $L$ of nodes with whom the node is connected. For a social network, the higher the density, the lower intermediate nodes are needed to connect two nodes. We redefine $L$ as follows:
\begin{equation}
  \label{eq:myerson-depth}
  L=\begin{cases}
    6, &\text{ if } D\leq 0.2,\\
    2, &\text{ if } 0.2<D\leq 0.3,\\
    1, &\text{ if } D>0.3,
  \end{cases}
\end{equation}
where $D$ is the density of the network.
 
The algorithm for calculation of Myerson values in weighted graphs is presented in Algorithm~\ref{alg:RefinedMyersonValue}. We use the same characteristic function as given by equation~\ref{eq:myerson-mazalov}, but we cutoff the maximal depth to approximate the Myerson value given by formula \eqref{eq:myerson-depth}.

\begin{algorithm}[!htbp]
\caption{Calculation of the approximated Myerson value for weighted graph}
\label{alg:RefinedMyersonValue}
\begin{algorithmic}[1]
  \Require A weighted graph $G(V,E,W)$, discount factor $r$ (default $0.5$), boolean $weight$ for considering edge weights (default True),  boolean $approximate$ for approximation (default True), boolean $scale$ for scaling (default False)
  \Ensure Components of Myerson value $Y_i(v,g)$, $i \in V$
  \State $Y_i(v,g) \gets 0$ for each $i \in V$ \Comment{Initialize Myerson values}

  \If {$approximate$}
    \State $L \gets \begin{cases} 
    1, & \text{if } density(G) > 0.3, \\
    2, & \text{if } density(G) > 0.2, \\
    6, & \text{otherwise}
    \end{cases}$ \Comment{Adjust $L$ based on $density(G)$.}
    \Else
    \State $L \gets |V|-1$ \Comment{Without approximation }
    \EndIf

  \ForAll{$i \in V$}
  \State $l2c\gets$ Initialize a length->count map for paths through $i$ 
    \ForAll{pairs $(start, end)$ in $V \times V$}
      \ForAll{$path$ in all simple paths from $start$ to $end$ with $length\leq L$}
          \If{$node \in path$}
              \State $length \gets \text{len}(path) - 1$
              \If{$weight$}
                \State $l2c[length] \gets l2c[length] + \min\limits_{(u,v) \in path} w(u,v)$
              \Else
                \State $l2c[length] \gets l2c[length] + 1$
              \EndIf
          \EndIf
      \EndFor
  \EndFor

  \ForAll{$(length, count) \in l2c$}
      \State $count \gets count/2$ \Comment{Correct for double counting}
      \State $Y_i(v,g) \gets Y_i(v,g) + \left(count \cdot \frac{r^{length}}{length + 1}\right)$
  \EndFor
  \EndFor
  \If {$scale$}
    \State Define $\xi \gets \frac{v(N)}{\sum_{i\in V}Y_i(v,g)}$ \Comment{For more accurate results}
    \State \Return $Y_i(v,g) \gets \xi \cdot Y_i(v,g)$ for all $i \in V$
  \Else
  \State \Return $Y_i(v,g)$ for all $i \in V$
  \EndIf
\end{algorithmic}
\end{algorithm}

Similarly to equation \eqref{eq:shapley-value-approximation-factor}, we can also define the scaling factor $\xi$ for the approximated Myerson value as follows:
\begin{equation}
    \label{eq:myerson-value-approximation-factor}
    \xi = \frac{v(N)}{\sum_{i\in V}Y_i(v,g)}.
\end{equation}

In both formulae \eqref{eq:shapley-value-approximation-factor} and \eqref{eq:myerson-value-approximation-factor}, $v(N)$ is used, and to calculate it for the Myerson value we need to count the number of paths for all lengths, i.e. $a_1, a_2, \ldots, a_L$. It is much more computationally expensive than in the case of the Shapley value. But after the $\xi$ scaling, we obtain  a more accurate approximation which will be shown in  Section~\ref{sec:centrality-experiments-description}.

\section{Experiments}

Network structure has a huge impact on key performance indicators (KPIs) of opinion dynamics realized on this network. Therefore, we define several  characteristics of a network which, in our opinion, have most significant correlation with KPIs of opinion dynamics. Our experiments are organized as follows: in Section \ref{sec:centrality-experiments-description} we provide the series of experiments in which we calculated the centralities based on approximated Shapley and Myerson values (realizations of Algorithms \ref{alg:shapley-approximation} and \ref{alg:RefinedMyersonValue}) for the graphs with varying density. We analyze correlation between KPIs of opinion dynamics and network properties in Section 
\ref{sec:experiments-description}.

\subsection{Centralities based on the Shapley and Myerson values}
\label{sec:centrality-experiments-description}

Due to the computational complexity of the Shapley value and Myerson value, we run our experiments on networks composed by 20 nodes for a given density. We create a network as follows: randomly and repeatedly take two different nodes from the set of nodes and add a connection between them until the density reaches a desired value. We designed the following experiments to evaluate the performance of the proposed centrality measures:
\begin{enumerate}
  \item [1.] {\bf Shapley-value based centrality:} The density of a network takes the values: $0.1, 0.2,\ldots, 1.0$. For each weighted or unweighted graph, we calculate the exact Shapley value and an approximated Shapley value. We compare (i) these two values and (ii)  computation time for these two methods. 
  \item [2.] {\bf Myerson-value based centrality:}  The density of a network takes the values: $0.1, 0.11,\ldots, 0.2$.\footnote{We limit the density to the set $\{0.1, 0.11,\ldots, 0.2\}$ because even a small increase in density significantly increases computation time. This makes the realization of the series of experiments quite complicated.} For each weighted or unweighted graph, we calculate the exact Myerson value and an approximated Myerson value. We compare: (i) these two values and (ii) computation time for these two methods. We also make the series of experiments on networks composed by 10 nodes with density from $0.05$ to $1.0$ to show efficiency of our algorithm to approximate the Myerson value.
  \item [3.] {\bf Comparison with classical centrality measures:} Based on the real social network dataset ``Zachary's karate club'', we created a two-layer network by adding different internal network structures. We reduce a two-layer network to a one-layer weighted network using opinion dynamics parameters by the method described in Section~\ref{sec:simplification}. The most important nodes in the network are nodes 0 (instructor --- Mr Hi) and  33 (manager --- John A). We define the coefficient of accuracy or simply accuracy of  centrality measures as follows:
\begin{equation}
  \label{eq:accuracy}
  Ac = \frac{|\text{Top 2 nodes according to a centrality measure} \cap \{0, 33\}|}{2}\cdot 100\%.
\end{equation}
The meaning of $Ac$ is the percentage of important nodes (0 and 33) in the top two nodes identified by considered centrality measures.
We also compare the accuracy $Ac$ of the proposed centrality measures with  the accuracy $Ac$ of classical centrality measures (betweenness and closeness centralities).
\end{enumerate}

The results of this part of experiments are presented in Tables \ref{tab:shapley-value-results-with-xi}--\ref{tab:centralities-benchmark}. In Table~\ref{tab:shapley-value-results-with-xi} we provide the results for the graphs with 20 nodes and different density (from 0.1 to 1.0), weighted and unweighted graphs. We compare the computation time for the exact Shapley value (column ``SV time'') and for an approximated Shapley value (column ``ASV time''). In Table \ref{tab:shapley-value-results-with-xi} we also present a root mean square error for an approximated Shapley value (column ``RMSE ASV'') and for the ratio of the approximated Shapley value component to the sum of all its components (column ``RMSE ratio ASV''). The lower the value of RMSE, the more accurate approximation we obtain. We can make the following conclusions based on the results from Table \ref{tab:shapley-value-results-with-xi}:
\begin{enumerate}
	\item The computation time of an approximated Shapley value is much faster than the time of the exact Shapley value (see columns ``SV time'' and ``ASV time'').
	\item The root mean square error of the approximated Shapley value is very small (see column ``RMSE ASV''), and root mean square error of the ratio is even smaller than the first one (see column ``RMSE ratio ASV''). The ratio here is referred to the normalized values. The RMSE of ratio is the RMSE between the exact normalized Shapley values and the approximated normalized Shapley values. RMSE increases with an increase of the graph density because of the sampling procedure. 
	\item The weighted graph introduces greater uncertainty in the sampling process when graph density is high ($\geq 0.7$), leading to a slightly higher RMSE. However, even in the worst case (see  graph ``20-1.0'' in  Table~\ref{tab:shapley-value-results-with-xi}), our algorithm still demonstrates a high accuracy. 
\end{enumerate}


\begin{table}[!htbp]
 \centering
 \caption{Results on the Shapley value when $\xi$ scaling factor is applied}
 \label{tab:shapley-value-results-with-xi}
 {\scriptsize
  \begin{filecontents*}{shapley_new_summary.csv}
graph,sv_time,asv_time,asv_scaled_time,rmse_sv,rmse_asv,rmse_asv_scaled,rmse_ratio_sv,rmse_ratio_asv,rmse_ratio_asv_scaled,weighted
20-0.1,715.6827371120453,0.002202749252319336,0.002156972885131836,0,2.6008234739935213e-22,2.6008234739935213e-22,0,5.467774254096536e-27,5.467774254096536e-27,True
20-0.2,870.5314631462097,0.01908087730407715,0.0261838436126709,0,3.106206019138013e-22,3.106206019138013e-22,0,5.6471829015634315e-27,5.6471829015634315e-27,True
20-0.3,1026.2327508926392,0.14658713340759277,0.13925385475158691,0,1.6390804057082854e-22,1.6390804057082854e-22,0,1.1628302691164582e-27,1.1628302691164582e-27,True
20-0.4,1172.074863910675,0.9346621036529541,0.9122960567474365,0,1.7166791970121652e-22,1.7166791970121652e-22,0,6.09404983180617e-28,6.09404983180617e-28,True
20-0.5,1318.2130370140076,2.225461006164551,2.1904470920562744,0,7.076960568326177e-23,7.076960568326177e-23,0,1.9693867436317972e-28,1.9693867436317972e-28,True
20-0.6,1460.767962217331,19.05217695236206,19.02755379676819,0,7.65989750767161e-23,7.65989750767161e-23,0,1.8415858833924533e-28,1.8415858833924533e-28,True
20-0.7,1579.210541009903,47.19689917564392,45.88863015174866,0,0.3791468370647083,0.03326053568498709,0,6.886517045514865e-07,6.041175174002384e-08,True
20-0.8,1689.9426691532135,173.42222499847412,174.28198909759521,0,12.769216428252223,1.9745802350922756,0,1.867038307814887e-05,2.887112894895319e-06,True
20-0.9,1789.2473990917206,250.35410523414612,250.21350598335266,0,49.38505790369246,6.865942445141016,0,5.673254676818809e-05,7.887454574544648e-06,True
20-1.0,1893.6711218357086,320.57276678085327,310.47058725357056,0,12.92038077176903,9.940871457666892,0,1.1741520383721889e-05,9.033862617051856e-06,True
20-0.1,718.8998630046844,0.003376007080078125,0.003408193588256836,0,4.4888915658949506e-23,4.4888915658949506e-23,0,1.1014960206424511e-26,1.1014960206424511e-26,False
20-0.2,878.5167498588562,0.01635289192199707,0.023921966552734375,0,1.054823724120434e-22,1.054823724120434e-22,0,7.844885060196077e-27,7.844885060196077e-27,False
20-0.3,1017.342572927475,0.09565186500549316,0.09852004051208496,0,1.0520297574636508e-22,1.0520297574636508e-22,0,6.632006318550484e-27,6.632006318550484e-27,False
20-0.4,1165.6817750930786,0.5467593669891357,0.4995851516723633,0,1.5136957402761854e-22,1.5136957402761854e-22,0,1.5862361844707587e-26,1.5862361844707587e-26,False
20-0.5,1320.8524091243744,2.9176790714263916,2.9878122806549072,0,1.0761448732772923e-22,1.0761448732772923e-22,0,8.595796146476193e-27,8.595796146476193e-27,False
20-0.6,1456.7016048431396,25.40595507621765,24.700889110565186,0,1.3657546702033134e-22,1.3657546702033134e-22,0,9.429935340207176e-27,9.429935340207176e-27,False
20-0.7,1573.5688781738281,62.25633096694946,60.69458889961243,0,0.09771247463832414,0.00037632484806444157,0,5.523911732587938e-06,2.127451229739963e-08,False
20-0.8,1682.8255372047424,172.952312707901,173.47865414619446,0,0.48514905433122624,0.0017684380356166145,0,2.0998487462316525e-05,7.654250499930852e-08,False
20-0.9,1782.0388910770416,251.51380801200867,251.0688030719757,0,0.659640053325793,0.002013491812518549,0,2.2558737845017916e-05,6.885851415954619e-08,False
20-1.0,1885.327553987503,319.1029267311096,309.4951560497284,0,4.1167183519739595e-21,4.1167183519739595e-21,0,8.613020551104498e-34,8.613020551104498e-34,False
\end{filecontents*}
\pgfplotstabletypeset[
      columns={graph,weighted,sv_time,asv_time,rmse_asv_scaled,rmse_ratio_asv_scaled},
      columns/graph/.style={
          column name=Graph,
          string type,
      },
      columns/sv_time/.style={
          column name={SV time},
          sci, precision=1, fixed zerofill
      },
      columns/asv_time/.style={
          column name={ASV time},
          sci, precision=1, fixed zerofill
      },
      columns/rmse_asv_scaled/.style={
          column name={RMSE ASV},
          sci, precision=1, fixed zerofill
      },
      columns/rmse_ratio_asv_scaled/.style={
          column name={RMSE ratio ASV},
          sci, precision=1, fixed zerofill
      },
      columns/weighted/.style={
          column name=Weighted,
          string type,
      },
      every last column/.style={column type/.add={}{|}},
      every row no 9/.style={after row=\midrule},
  ]{shapley_new_summary.csv}
}
\end{table}

In Tables~\ref{tab:myerson-value-results} and \ref{tab:myerson-value-results-with-xi} we present the results on computation of the Myerson value (exact and approximated) without and with scaling factor $\xi$ defined by \eqref{eq:myerson-value-approximation-factor}, respectively. In both tables we provide the results for the graphs with 20 nodes and different density (from 0.1 to 0.2), weighted and unweighted graphs. We compare the computation time for the exact Myerson value (column ``MV time'') and for an approximated Myerson value (column ``AMV time''). In these two tables   we also present a root mean square error for an approximated Myerson value (column ``RMSE AMV'') and for the ratio of the approximated Myerson value component to the sum of all its components (column ``RMSE ratio AMV''). Conclusions from Tables ~\ref{tab:myerson-value-results} and \ref{tab:myerson-value-results-with-xi} are
\begin{enumerate}
	\item The computational complexity of the exact Myerson-value based centrality grows rapidly with the increase of the network density (column ``MV time'').
	\item The root mean square error of an approximated Myerson value grows with an increase of the network  density (see column ``RMSE AMV''), but the root mean square error of the ratio is very small. The RMSE of the ratio means the RMSE between the normalized approximated  and normalized exact Myerson values. Therefore, we can recommend to use an approximation of the Myerson value as an approximated centrality measure.
	\item If we estimate the effect of scaling  factor $\xi$ on the results, we can compare ``AMV time''  in Tables~\ref{tab:myerson-value-results} and \ref{tab:myerson-value-results-with-xi}, and conclude that without $\xi$ the computation time is much smaller, but the RMSE AMV is higher. While the RMSE ratio AMV between Tables~\ref{tab:myerson-value-results} and \ref{tab:myerson-value-results-with-xi} are almost the same.
	\item Comparing  Tables~\ref{tab:myerson-value-results} and \ref{tab:myerson-value-results-with-xi}, we cannot see a significant difference in the results referred to weighted and unweighted graphs.
\end{enumerate}

\begin{table}[!htbp]
 \centering
 \caption{Results on the Myerson value without $\xi$ scaling}
 \label{tab:myerson-value-results}
  {\scriptsize
  \begin{filecontents*}{myerson_summary.csv}
graph,mv_time,amv_time,rmse_mv,rmse_amv,rmse_ratio_mv,rmse_ratio_amv,weighted
20-0.1,0.10316681861877441,0.16190409660339355,0,0.0004366003922504856,0,1.4771208207361577e-08,True
20-0.11,0.1907520294189453,0.1670830249786377,0,0.0008750226074332085,0,1.6987841400311618e-08,True
20-0.12,0.4170560836791992,0.22285890579223633,0,0.15440462856265255,0,5.277177894146352e-07,True
20-0.13,1.0215330123901367,0.35552000999450684,0,0.5279639344450641,0,8.149795154812228e-07,True
20-0.14,2.210071086883545,0.6152961254119873,0,2.136497370912902,0,1.9343592767136186e-06,True
20-0.15,8.775124311447144,0.9243769645690918,0,18.906849586383203,0,4.297079350052703e-06,True
20-0.16,17.98325777053833,1.4792709350585938,0,127.29743562922239,0,1.2526003454294142e-05,True
20-0.17,28.74915909767151,2.1431632041931152,0,405.65651814230057,0,3.2380762424255435e-05,True
20-0.18,63.36852312088013,2.7690258026123047,0,1546.7709534971395,0,4.7018616044183415e-05,True
20-0.19,144.5796091556549,3.579231023788452,0,4818.114936977139,0,8.656355938172019e-05,True
20-0.2,294.11523818969727,4.445711851119995,0,13256.499090683328,0,0.00013455993176979977,True
20-0.1,0.3295018672943115,0.196303129196167,0,0.011184135176428295,0,1.4321184307985638e-06,False
20-0.11,0.3449368476867676,0.21645188331604004,0,0.017296756969918083,0,1.4100873534161024e-06,False
20-0.12,0.48670101165771484,0.21749186515808105,0,0.03427762886869244,0,1.519863312693138e-06,False
20-0.13,1.366335153579712,0.3785221576690674,0,0.3949606714945735,0,5.729475484846506e-06,False
20-0.14,3.6937012672424316,0.5808889865875244,0,2.7123371549032056,0,1.5349264225825544e-05,False
20-0.15,7.652312994003296,1.0546510219573975,0,13.777553444317148,0,2.7493083759844427e-05,False
20-0.16,17.05877709388733,1.3823411464691162,0,59.35931839553511,0,4.192227847917749e-05,False
20-0.17,42.796765089035034,1.6157870292663574,0,297.08142879881683,0,6.294927301702355e-05,False
20-0.18,82.54166293144226,1.884131908416748,0,803.8331284710878,0,6.731490215726592e-05,False
20-0.19,168.71626496315002,2.5634119510650635,0,2722.5537421960757,0,7.60139007460394e-05,False
20-0.2,366.35711002349854,3.568497896194458,0,8745.385979038432,0,0.00013125599806120791,False
\end{filecontents*}
\pgfplotstabletypeset[
      columns={graph,weighted,mv_time,amv_time,rmse_amv,rmse_ratio_amv},
      columns/graph/.style={
          column name=Graph,
          string type,
      },
      columns/mv_time/.style={
          column name={MV time},
          sci, precision=2, fixed zerofill
      },
      columns/amv_time/.style={
          column name={AMV time},
          sci, precision=2, fixed zerofill
      },
      columns/rmse_amv/.style={
          column name={RMSE AMV},
          sci, precision=2, fixed zerofill
      },
      columns/rmse_ratio_amv/.style={
          column name={RMSE ratio AMV},
          sci, precision=2, fixed zerofill
      },
      columns/weighted/.style={
          column name=Weighted,
          string type,
      },
      every last column/.style={column type/.add={}{|}},
      every row no 10/.style={after row=\midrule},
  ]{myerson_summary.csv}
  }
\end{table}

\begin{table}[!htbp]
 \centering
 \caption{Results on the Myerson value when $\xi$ scaling is applied}
 \label{tab:myerson-value-results-with-xi}
  {\scriptsize
  \begin{filecontents*}{myerson_new_summary.csv}
graph,mv_time,amv_time,rmse_mv,rmse_amv,rmse_ratio_mv,rmse_ratio_amv,weighted
20-0.1,0.1232290267944336,0.10995197296142578,0,0.00019803916887604363,0,1.4771208207362045e-08,True
20-0.11,0.15680909156799316,0.21091914176940918,0,0.0003207385579957342,0,1.6987841400311618e-08,True
20-0.12,0.45294189453125,0.22144007682800293,0,0.015451800328354297,0,5.277177894146396e-07,True
20-0.13,0.909512996673584,0.43328261375427246,0,0.031513953528544573,0,8.149795154812214e-07,True
20-0.14,2.0947840213775635,0.6696600914001465,0,0.10848197395706499,0,1.934359276713615e-06,True
20-0.15,7.825020790100098,1.2386012077331543,0,0.4814701077067506,0,4.297079350052705e-06,True
20-0.16,16.518309116363525,2.2392046451568604,0,3.85239684856348,0,1.2526003454294142e-05,True
20-0.17,26.281731843948364,3.280039072036743,0,23.592171124526292,0,3.2380762424255435e-05,True
20-0.18,59.10757398605347,5.906550884246826,0,61.546016539684935,0,4.701861604418362e-05,True
20-0.19,141.4668481349945,11.522468090057373,0,206.7780409441743,0,8.656355938172019e-05,True
20-0.2,291.59159874916077,20.614261150360107,0,647.9624080610611,0,0.0001345599317697998,True
20-0.1,0.2749011516571045,0.24292278289794922,0,0.0015931701796828934,0,1.4321184307985627e-06,False
20-0.11,0.36444902420043945,0.1936960220336914,0,0.0018144092679479122,0,1.4100873534161012e-06,False
20-0.12,0.4558529853820801,0.24716711044311523,0,0.0026452523363911145,0,1.5198633126931334e-06,False
20-0.13,1.339980125427246,0.4131131172180176,0,0.020097730783168666,0,5.729475484846469e-06,False
20-0.14,3.329132080078125,0.6969528198242188,0,0.11823156594953178,0,1.5349264225825554e-05,False
20-0.15,7.174816131591797,1.2909648418426514,0,0.5637592047698523,0,2.749308375984435e-05,False
20-0.16,15.50442385673523,2.008274793624878,0,1.919238632890965,0,4.1922278479177485e-05,False
20-0.17,39.34808826446533,3.304759979248047,0,7.443539144189997,0,6.29492730170237e-05,False
20-0.18,77.8272430896759,5.57743501663208,0,15.954148620216868,0,6.731490215726592e-05,False
20-0.19,163.83924293518066,10.606647968292236,0,45.77776333247767,0,7.601390074603945e-05,False
20-0.2,343.01345014572144,19.882071256637573,0,205.83484201492857,0,0.00013125599806120802,False
\end{filecontents*}
\pgfplotstabletypeset[
      columns={graph,weighted,mv_time,amv_time,rmse_amv,rmse_ratio_amv},
      columns/graph/.style={
          column name=Graph,
          string type,
      },
      columns/mv_time/.style={
          column name={MV time},
          sci, precision=2, fixed zerofill
      },
      columns/amv_time/.style={
          column name={AMV time},
          sci, precision=2, fixed zerofill
      },
      columns/rmse_amv/.style={
          column name={RMSE AMV},
          sci, precision=2, fixed zerofill
      },
      columns/rmse_ratio_amv/.style={
          column name={RMSE ratio AMV},
          sci, precision=2, fixed zerofill
      },
      columns/weighted/.style={
          column name=Weighted,
          string type,
      },
      every last column/.style={column type/.add={}{|}},
      every row no 10/.style={after row=\midrule},
  ]{myerson_new_summary.csv}
  }
\end{table}

Next, we demonstrate the efficiency of our algorithm to approximate the Myerson value in high-density graphs. To do this, we examine graphs with $10$ nodes and with density from $0.05$ to $1.0$ with step $0.05$, weighted and unweighted graphs. The results of our experiments are presented in Tables~\ref{tab:myerson-value-results-10} and \ref{tab:myerson-value-results-with-xi-10} for the cases without and with scaling factor, respectively.

\begin{table}[!htbp]
  \centering
  \caption{Results on the Myerson value without $\xi$ scaling for networks with 10 nodes}
  \label{tab:myerson-value-results-10}
   {\scriptsize
   \begin{filecontents*}{myerson_summary_10.csv}
graph,mv_time,amv_time,rmse_mv,rmse_amv,rmse_ratio_mv,rmse_ratio_amv,weighted
10-0.05,0.0012400150299072266,0.015869855880737305,0,0.0,0,0.0,True
10-0.1,0.0017867088317871094,0.0015790462493896484,0,0.0,0,0.0,True
10-0.15,0.01267099380493164,0.0031690597534179688,0,0.0,0,0.0,True
10-0.2,0.007330656051635742,0.010500669479370117,0,9.650505554713072e-06,0,7.63439494484995e-10,True
10-0.25,0.014878988265991211,0.002892017364501953,0,32.6572597199336,0,0.00016443798102638816,True
10-0.3,0.06692790985107422,0.0033538341522216797,0,279.23472965662785,0,0.0001911690325002184,True
10-0.35,0.11743283271789551,0.0017788410186767578,0,1854.492299423709,0,0.0005121311636951157,True
10-0.4,0.2601039409637451,0.001196146011352539,0,4578.215794600516,0,0.0004076088950919469,True
10-0.45,0.5531599521636963,0.0012061595916748047,0,10508.865999301992,0,0.000625677418868713,True
10-0.5,1.2119150161743164,0.0013661384582519531,0,15557.624394628692,0,0.0005530410080385413,True
10-0.55,2.5555267333984375,0.0012629032135009766,0,39219.07865639237,0,0.00044902146947733767,True
10-0.6,5.701395034790039,0.0016188621520996094,0,110892.99469570685,0,0.0006174686571174432,True
10-0.65,10.488188028335571,0.001325845718383789,0,231415.52512284095,0,0.0007474540744370548,True
10-0.7,19.91686725616455,0.0013303756713867188,0,519110.6146890814,0,0.0006844059467056281,True
10-0.75,35.46976900100708,0.006443023681640625,0,1003751.5064183117,0,0.0005961079583184776,True
10-0.8,73.57436490058899,0.0013790130615234375,0,3417946.095826972,0,0.0005904844664054744,True
10-0.85,115.38576602935791,0.0013849735260009766,0,8106031.580318517,0,0.0003760272920692568,True
10-0.9,173.1053249835968,0.001386880874633789,0,16304105.778097097,0,0.000392205113108905,True
10-0.95,248.39101815223694,0.0013659000396728516,0,27437718.545221668,0,0.0003697832398353691,True
10-1.0,433.0312030315399,0.0013737678527832031,0,73091096.4290595,0,0.0002793943159805221,True
10-0.05,0.0011518001556396484,0.0010569095611572266,0,0.0,0,0.0,False
10-0.1,0.0020852088928222656,0.001992940902709961,0,0.0,0,0.0,False
10-0.15,0.0033981800079345703,0.0028362274169921875,0,0.0,0,0.0,False
10-0.2,0.007573843002319336,0.01084280014038086,0,0.0,0,0.0,False
10-0.25,0.031358957290649414,0.00837087631225586,0,2.5575256843116434,0,0.00015672989421740348,False
10-0.3,0.031219959259033203,0.0031096935272216797,0,8.849855659081207,0,9.039598216430807e-05,False
10-0.35,0.08494114875793457,0.0011327266693115234,0,59.81606997178687,0,0.00015320904864402532,False
10-0.4,0.21819591522216797,0.0013272762298583984,0,228.13602971229463,0,0.00012947975195756585,False
10-0.45,0.47412610054016113,0.0011649131774902344,0,622.7845276366608,0,9.579275711321291e-05,False
10-0.5,0.9613497257232666,0.0011680126190185547,0,1813.1932119722644,0,8.713092218795374e-05,False
10-0.55,1.9477379322052002,0.0012059211730957031,0,4914.8927451912405,0,7.402509614291985e-05,False
10-0.6,4.154980182647705,0.0013041496276855469,0,19695.757358614723,0,4.9768002408546285e-05,False
10-0.65,7.106881856918335,0.0012407302856445312,0,47944.05390414132,0,5.334768466475176e-05,False
10-0.7,11.422459125518799,0.001331329345703125,0,106956.1938486338,0,9.369425126879216e-05,False
10-0.75,18.57507300376892,0.0035102367401123047,0,245659.215221752,0,0.00015950941267691207,False
10-0.8,37.66133093833923,0.001390218734741211,0,853420.833223208,0,0.00010740939048779022,False
10-0.85,52.555420875549316,0.001291036605834961,0,1727248.512509386,0,0.00013468339267762398,False
10-0.9,74.5032889842987,0.0013051033020019531,0,3644994.671602051,0,9.854801284601043e-05,False
10-0.95,104.90536499023438,0.0013010501861572266,0,7735922.974459719,0,4.8535648440568956e-05,False
10-1.0,170.9519760608673,0.001275777816772461,0,21634704.083071303,0,0.0,False
\end{filecontents*}
\pgfplotstabletypeset[
       columns={graph,weighted,mv_time,amv_time,rmse_amv,rmse_ratio_amv},
       columns/graph/.style={
           column name=Graph,
           string type,
       },
       columns/mv_time/.style={
           column name={MV time},
           sci, precision=2, fixed zerofill
       },
       columns/amv_time/.style={
           column name={AMV time},
           sci, precision=2, fixed zerofill
       },
       columns/rmse_amv/.style={
           column name={RMSE AMV},
           sci, precision=2, fixed zerofill
       },
       columns/rmse_ratio_amv/.style={
           column name={RMSE ratio AMV},
           sci, precision=2, fixed zerofill
       },
       columns/weighted/.style={
           column name=Weighted,
           string type,
       },
       every last column/.style={column type/.add={}{|}},
       every row no 19/.style={after row=\midrule},
   ]{myerson_summary_10.csv}
   }
 \end{table}
 
 \begin{table}[!htbp]
  \centering
  \caption{Results on the Myerson value when $\xi$ scaling is applied for networks with 10 nodes}
  \label{tab:myerson-value-results-with-xi-10}
   {\scriptsize
   \begin{filecontents*}{myerson_new_summary_10.csv}
graph,mv_time,amv_time,rmse_mv,rmse_amv,rmse_ratio_mv,rmse_ratio_amv,weighted
10-0.05,0.0012390613555908203,0.001219034194946289,0,0.0,0,0.0,True
10-0.1,0.0030829906463623047,0.0019903182983398438,0,6.236493043901527e-32,0,0.0,True
10-0.15,0.00466609001159668,0.0037310123443603516,0,0.0,0,0.0,True
10-0.2,0.01203608512878418,0.017444849014282227,0,3.251948249971347e-06,0,7.63439494484995e-10,True
10-0.25,0.01459813117980957,0.0044820308685302734,0,1.6712095369415785,0,0.00016443798102638824,True
10-0.3,0.04972195625305176,0.00824427604675293,0,6.054873056938856,0,0.00019116903250021808,True
10-0.35,0.10508012771606445,0.016850948333740234,0,35.02277383577499,0,0.0005121311636951152,True
10-0.4,0.2566196918487549,0.02626800537109375,0,64.10636665009497,0,0.000407608895091947,True
10-0.45,0.5516719818115234,0.07068490982055664,0,214.0195235575375,0,0.000625677418868713,True
10-0.5,1.1765906810760498,0.12743306159973145,0,291.11678295965004,0,0.0005530410080385412,True
10-0.55,2.5008749961853027,0.3797457218170166,0,578.3451290824478,0,0.00044902146947733767,True
10-0.6,5.807449817657471,0.576289176940918,0,2148.6756733739644,0,0.0006174686571174435,True
10-0.65,10.647480964660645,1.2367579936981201,0,5357.971346898657,0,0.0007474540744370548,True
10-0.7,19.798518657684326,2.1359896659851074,0,11215.663111830943,0,0.0006844059467056281,True
10-0.75,35.73496985435486,3.659411907196045,0,18988.46737225138,0,0.0005961079583184773,True
10-0.8,74.2675130367279,6.771676063537598,0,64107.0536538134,0,0.0005904844664054743,True
10-0.85,115.64395809173584,11.142996311187744,0,97121.90743291365,0,0.00037602729206925693,True
10-0.9,175.37625193595886,17.05469584465027,0,203490.98462711423,0,0.0003922051131089048,True
10-0.95,247.81292009353638,24.542087078094482,0,322332.8081187206,0,0.0003697832398353692,True
10-1.0,432.5113310813904,43.329052209854126,0,648869.0313266467,0,0.0002793943159805221,True
10-0.05,0.001233816146850586,0.0012850761413574219,0,0.0,0,0.0,False
10-0.1,0.0022759437561035156,0.002547025680541992,0,0.0,0,0.0,False
10-0.15,0.004912853240966797,0.017326831817626953,0,0.0,0,0.0,False
10-0.2,0.021509885787963867,0.00795602798461914,0,2.1632835246033423e-31,0,2.649291478610512e-33,False
10-0.25,0.016412019729614258,0.009183883666992188,0,0.05345974041225476,0,0.00015672989421740348,False
10-0.3,0.04557490348815918,0.0064830780029296875,0,0.0723013371993383,0,9.039598216430815e-05,False
10-0.35,0.06445789337158203,0.00683903694152832,0,0.3233649804044364,0,0.00015320904864402532,False
10-0.4,0.2428591251373291,0.0199282169342041,0,1.0175225183571168,0,0.00012947975195756617,False
10-0.45,0.4541809558868408,0.03733706474304199,0,2.0019073475789786,0,9.579275711321291e-05,False
10-0.5,0.9118270874023438,0.08233499526977539,0,5.101088140968938,0,8.713092218795371e-05,False
10-0.55,1.859415054321289,0.21135687828063965,0,11.471772347951214,0,7.402509614291975e-05,False
10-0.6,4.287614822387695,0.36111998558044434,0,30.30585004561893,0,4.97680024085464e-05,False
10-0.65,7.357494115829468,0.6801180839538574,0,78.2839603988513,0,5.334768466475176e-05,False
10-0.7,11.432630777359009,1.506514310836792,0,305.07326806373646,0,9.369425126879216e-05,False
10-0.75,19.388360023498535,2.025787115097046,0,1233.3034373281869,0,0.00015950941267691207,False
10-0.8,37.35586190223694,3.4067931175231934,0,2900.8427190082907,0,0.00010740939048779045,False
10-0.85,52.712162017822266,5.042814016342163,0,7345.268844534856,0,0.00013468339267762414,False
10-0.9,74.49852991104126,6.499423980712891,0,11346.581134241897,0,9.854801284601022e-05,False
10-0.95,104.90217804908752,9.533251762390137,0,11886.333008139403,0,4.8535648440569024e-05,False
10-1.0,171.19001698493958,16.326950788497925,0,0.0,0,0.0,False
\end{filecontents*}
\pgfplotstabletypeset[
       columns={graph,weighted,mv_time,amv_time,rmse_amv,rmse_ratio_amv},
       columns/graph/.style={
           column name=Graph,
           string type,
       },
       columns/mv_time/.style={
           column name={MV time},
           sci, precision=2, fixed zerofill
       },
       columns/amv_time/.style={
           column name={AMV time},
           sci, precision=2, fixed zerofill
       },
       columns/rmse_amv/.style={
           column name={RMSE AMV},
           sci, precision=2, fixed zerofill
       },
       columns/rmse_ratio_amv/.style={
           column name={RMSE ratio AMV},
           sci, precision=2, fixed zerofill
       },
       columns/weighted/.style={
           column name=Weighted,
           string type,
       },
       every last column/.style={column type/.add={}{|}},
       every row no 19/.style={after row=\midrule},
   ]{myerson_new_summary_10.csv}
   }
 \end{table}

By comparing the pairs of Tables~\ref{tab:myerson-value-results} and ~\ref{tab:myerson-value-results-with-xi},  and Tables~\ref{tab:myerson-value-results-10} and \ref{tab:myerson-value-results-with-xi-10}, we can conclude that the scaling factor $\xi$ has a very limited effect on reducing the error of an approximated Myerson value and moreover using $\xi$ significantly increases computation time.  But the RMSE of ratio is very small. Therefore, we could recommend to use the Myerson-value based centrality approximation without scaling factor $\xi$ and better its ratio as an approximation of the Myerson value.

In Table~\ref{tab:shapley-value-results-with-xi}, the results for the weighted graph with density $1.0$ show the highest RMSE, and in Table~\ref{tab:myerson-value-results-with-xi}
the results for the weighted graph with density $0.2$  also show the highest RMSE. Therefore, we decide to carefully examine the high-density case and examine the Shapley value for the weighted graph with density $1.0$ and the Myerson value for the weighted graph with density $0.2$. To make analysis, we show the components of the exact and approximated Shapley values in Table~\ref{tab:shapley-value-graph-1.0} for the former graph and the exact and approximated Myerson values in  Table~\ref{tab:myerson-value-graph-0.2} for the latter graph. In these tables we provide a relative error which is calculated by taking the difference between the approximated value and the exact value  dividing this difference by the exact value, that is, $\frac{\text{approximated value}-\text{exact value}}{\text{exact value}}$.

Analyzing Table~\ref{tab:shapley-value-graph-1.0}, we can say that our algorithm provides a very accurate approximation for the Shapley value such that only for two nodes (Nodes 5 and 19) the absolute relative error is  larger than 5\%.

\begin{table}[!htbp]
 \centering
 \caption{Exact Shapley value vs an approximated Shapley value for ``graph-1.0''}
 \label{tab:shapley-value-graph-1.0}
  {\scriptsize
  \begin{filecontents*}{shapley_new_weighted_result-20-1.0.csv}
node,real_sv,real_sv_ratio,approx_sv,approx_sv_error,approx_sv_ratio,approx_sv_ratio_error,approx_sv_scaled,approx_sv_scaled_error,approx_sv_scaled_ratio,approx_sv_scaled_ratio_error
0.0,52.0000000000021,0.04957102001906515,52.947384521260325,0.9473845212582219,0.050474151116549404,0.0009031310974842513,53.35788035226238,1.357880352260274,0.05086547221378682,0.0012944521947216708
1.0,46.500000000003695,0.04432793136320422,45.24594677271337,1.2540532272903278,0.043132456408687674,0.0011954749545165441,46.50895839659884,0.008958396595147633,0.044336471302763435,8.539939559217002e-06
2.0,50.50000000000503,0.04814108674928727,48.29443254817987,2.2055674518251607,0.046038543897216275,0.002102542852070996,49.05739447312481,1.4426055268802216,0.04676586699058609,0.0013752197587011822
3.0,53.0000000000006,0.05052430886558416,54.71229733863567,1.712297338635075,0.05215662282043439,0.00163231395485023,53.67643486182812,0.6764348618275235,0.05116914667476465,0.0006448378091804957
4.0,50.500000000002544,0.048141086749284905,53.268277760783114,2.7682777607805704,0.05078005506271031,0.002638968313425402,50.01305800182205,0.486941998180491,0.04767689037351959,0.00046419637576531475
5.0,50.00000000000573,0.04766444232602771,50.380238605078006,0.3802386050722788,0.04802691954726216,0.00036247722123444737,54.313543880959614,4.313543880953887,0.05177649559672032,0.004112053270692605
6.0,50.00000000000588,0.047664442326027866,52.145151422453345,2.145151422447462,0.049709391251147136,0.0020449489251192707,51.287276040085025,1.287276040079142,0.04889158821743091,0.0012271458914030475
7.0,63.00000000000351,0.06005719733079139,63.69730804527378,0.6973080452702689,0.060721933312939735,0.0006647359821483484,62.595961129669,0.40403887033451014,0.059672031582143945,0.00038516574864744246
8.0,54.999999999996014,0.05243088655862068,51.824258182930556,3.1757418170654574,0.049403487304986234,0.003027399253634447,53.35788035226238,1.6421196477336366,0.05086547221378682,0.001565414344833857
9.0,43.99999999999913,0.041944709246898755,47.3317528296115,3.3317528296123697,0.04512083205873356,0.003176122811834804,44.4383540844215,0.43835408442236456,0.04236258730640753,0.0004178780595087764
10.0,49.50000000000465,0.04718779790276646,49.41755888650964,0.08244111349500827,0.047109207708779445,7.859019398701733e-05,50.331612511387796,0.831612511383149,0.04798056483449742,0.0007927669317309566
11.0,51.000000000005926,0.04861773117254835,50.701131844600795,0.2988681554051311,0.04833282349342306,0.0002849076791252847,51.60583054965077,0.6058305496448426,0.04919526267840874,0.0005775315058603944
12.0,64.00000000000799,0.0610104861773161,63.69730804527378,0.3026919547342075,0.060721933312939735,0.00028855286437636657,62.91451563923474,1.0854843607732434,0.059975706043121774,0.0010347801341943283
13.0,49.00000000000653,0.046711153479508036,47.49219944937289,1.5078005506336396,0.04527378403181401,0.0014373694476940285,49.05739447312481,0.057394473118279166,0.04676586699058609,5.471351107805278e-05
14.0,48.000000000003595,0.04575786463298479,48.77577240746405,0.7757724074604582,0.04649739981645763,0.0007395351834728406,47.9424536896447,0.05754631035889446,0.04570300637716368,5.48582558211122e-05
15.0,63.50000000000457,0.06053384175405262,61.13016212909147,2.369837870913102,0.05827470174365249,0.002259140010400125,63.073792894017615,0.4262071059869541,0.06012754327361069,0.0004062984804419287
16.0,53.99999999999861,0.05147759771210271,52.46604466197614,1.5339553380224658,0.050015295197308046,0.0014623025147946639,54.472821135742485,0.4728211357438781,0.05192833282720923,0.00045073511510652225
17.0,56.999999999994635,0.05433746425166026,57.92122973386356,0.9212297338689268,0.055215662282043436,0.0008781980303831788,57.02125721226845,0.021257212273816606,0.054357728515031886,2.0264263371629065e-05
18.0,53.00000000000058,0.05052430886558414,53.268277760783114,0.2682777607825315,0.05078005506271031,0.00025574619712616375,52.08366231399939,0.9163376860011923,0.04965077436987549,0.0008735344957086499
19.0,45.50000000000438,0.04337464251668442,44.283267054145,1.216732945859377,0.04221474457020496,0.0011598979464794643,41.88991800789554,3.6100819921088387,0.03993319161858488,0.003441450898099545
\end{filecontents*}
\pgfplotstabletypeset[
      columns={node,real_sv,real_sv_ratio,approx_sv_scaled,approx_sv_scaled_ratio,rel_error},
      create on use/rel_error/.style={
          create col/expr={
              (\thisrow{approx_sv_scaled} - \thisrow{real_sv}) / \thisrow{real_sv}
          }
      },
      columns/node/.style={
          column name=Nodes,
          fixed, precision=0, fixed zerofill
      },
      columns/real_sv/.style={
          column name={SV},
          fixed, precision=4, fixed zerofill
      },
      columns/real_sv_ratio/.style={
          column name={SV ratio},
          postproc cell content/.append code={\pgfkeysalso{@cell content=\pgfmathparse{##1*100}\pgfmathprintnumber[zerofill]{\pgfmathresult}\%}}
      },
      columns/approx_sv_scaled/.style={
          column name={ASV},
          fixed, precision=4, fixed zerofill
      },
      columns/approx_sv_scaled_ratio/.style={
          column name={ASV ratio},
          postproc cell content/.append code={\pgfkeysalso{@cell content=\pgfmathparse{##1*100}\pgfmathprintnumber[zerofill]{\pgfmathresult}\%}}
      },
      columns/rel_error/.style={
          column name={Rel. error ASV/SV},
          postproc cell content/.append code={\pgfkeysalso{@cell content=\pgfmathparse{##1*100}\pgfmathprintnumber[zerofill]{\pgfmathresult}\%}}
      },
      every last column/.style={column type/.add={}{|}},
  ]{shapley_new_weighted_result-20-1.0.csv}
  }
\end{table}

Analyzing the results in Table~\ref{tab:myerson-value-graph-0.2}, we can find that some nodes (Nodes 1, 2, 8, and 9) have a high relative error in approximation. The reason is that in our algorithm we count the number of paths not for all lengths ($length \leq L$). By increasing $L$ we can obtain a more accurate result, but it will increase the computational complexity.

\begin{table}[!htbp]
 \centering
 \caption{Exact Myerson value vs an approximated Myerson value for ``graph-0.2''}
 \label{tab:myerson-value-graph-0.2}
  {\scriptsize
  \begin{filecontents*}{myerson_new_weighted_result-20-0.2.csv}
node,real_mv,real_mv_ratio,approx_mv,approx_mv_error,approx_mv_ratio,approx_mv_ratio_error
0.0,143.2671554789471,0.06528743582566343,144.02222222075935,0.755066741812243,0.0656315228656089,0.0003440870399454704
1.0,68.70017561125817,0.031306954419786245,49.33695746582237,19.363218145435802,0.022483055754231804,0.008823898665554442
2.0,68.0766524475029,0.031022812332903885,81.63346607391703,13.55681362641414,0.03720070842273764,0.006177896089833757
3.0,104.67991895988342,0.04770307240680385,109.97629159177578,5.296372631892353,0.05011665134021321,0.0024135789334093632
4.0,164.08796143277047,0.07477556328938338,157.18532834497586,6.902633087794612,0.07163000481688536,0.0031455584724980196
5.0,114.63884012433556,0.052241394007755536,106.14993140418922,8.488908720146341,0.04837296316299076,0.003868430844764778
6.0,16.77053853961985,0.0076424038364336835,17.857771138466465,1.0872325988466152,0.008137860232475539,0.0004954563960418556
7.0,228.8443832212716,0.10428533276643305,251.43070181361063,22.586318592339012,0.11457801164810463,0.010292678881671583
8.0,33.768186133913105,0.015388302209230908,24.312802070897995,9.45538406301511,0.011079444549864493,0.004308857659366415
9.0,36.183189325174766,0.01648882916072355,28.18267092333121,8.000518401843557,0.012842959805767612,0.0036458693549559374
10.0,184.17398480059387,0.08392884728693054,192.6531648490396,8.479180048445727,0.08779284473572815,0.003863997448797618
11.0,79.79014728228503,0.03636070041872169,69.86023490587513,9.929912376409902,0.03183559824256667,0.004525102176155024
12.0,56.79061018913913,0.025879716155658105,53.00663307509619,3.7839771140429406,0.02415534212049998,0.0017243740351581237
13.0,48.86781113865808,0.022269263830847905,46.0495174888209,2.8182936498371802,0.0209849557479066,0.0012843080829413045
14.0,107.3196486400729,0.04890600814958681,93.90691342553325,13.412735214539651,0.04279376918847646,0.00611223896111035
15.0,63.40394062578923,0.028893438212421827,66.58614114508427,3.182200519295037,0.030343580162217127,0.0014501419497953
16.0,186.61460258781398,0.08504104691583655,200.20285243506672,13.58824984725274,0.09123326862164097,0.006192221705804413
17.0,229.37781345299672,0.1045284191312373,258.18789108105074,28.810077628054017,0.11765729076958259,0.013128871638345294
18.0,162.64728259958918,0.07411903998121638,152.48212175235116,10.165160847238013,0.06948673410306189,0.004632305878154497
19.0,96.40338070550408,0.04393141966242557,91.38261009145536,5.020770614048715,0.04164343370943963,0.0022879859529859395
\end{filecontents*}
\pgfplotstabletypeset[
      columns={node,real_mv,real_mv_ratio,approx_mv,approx_mv_ratio,rel_error},
      create on use/rel_error/.style={
          create col/expr={
              (\thisrow{approx_mv} - \thisrow{real_mv}) / \thisrow{real_mv}
          }
      },
      columns/node/.style={
          column name=Nodes,
          fixed, precision=0, fixed zerofill
      },
      columns/real_mv/.style={
          column name={MV},
          fixed, precision=4, fixed zerofill
      },
      columns/real_mv_ratio/.style={
          column name={MV ratio},
          postproc cell content/.append code={\pgfkeysalso{@cell content=\pgfmathparse{##1*100}\pgfmathprintnumber[zerofill]{\pgfmathresult}\%}}
      },
      columns/approx_mv/.style={
          column name={AMV},
          fixed, precision=4, fixed zerofill
      },
      columns/approx_mv_ratio/.style={
          column name={AMV ratio},
          postproc cell content/.append code={\pgfkeysalso{@cell content=\pgfmathparse{##1*100}\pgfmathprintnumber[zerofill]{\pgfmathresult}\%}}
      },
      columns/rel_error/.style={
          column name={Rel. error AMV/MV},
          postproc cell content/.append code={\pgfkeysalso{@cell content=\pgfmathparse{##1*100}\pgfmathprintnumber[zerofill]{\pgfmathresult}\%}}
      },
      every last column/.style={column type/.add={}{|}},
  ]{myerson_new_weighted_result-20-0.2.csv}
  }
\end{table}

The last item in analysis of this series of experiments is to compare the accuracy defined by \eqref{eq:accuracy} for the approximated Shapley and Myerson values, and for the classical centrality measures (betweenness and closeness). We summarize these results in  Table~\ref{tab:centralities-benchmark}. Both centrality measures proposed in this paper have a higher accuracy than the classical centrality measures. In particular, an approximated Shapley-value based centrality can identify the most important nodes with 100\% accuracy for all examined graphs. For the approximated Myerson value based centrality, it has 100\% accuracy for all cases except a case of a network ``karate-twoStar-34'', for which the accuracy is 50\%. The mistake in prediction of one of the most important nodes is that another node (Node 17) is the central one in the internal layer by construction. For the betweenness centrality, the accuracy is zero for high-density networks. Closeness centrality also has the  worst accuracy for these networks.

\begin{table}[!htbp]
 \centering
 \caption{Accuracy defined by \eqref{eq:accuracy} for the proposed in this paper and classical centrality measures}
 \label{tab:centralities-benchmark}
  \begin{filecontents*}{weighted_karate_summary.csv}
graph,bc_accuracy,cc_accuracy,sv_accuracy,mv_accuracy
karate-34,1.0,0.5,1.0,1.0
karate-empty-34,1.0,0.5,1.0,1.0
karate-karate-34,1.0,0.5,1.0,1.0
karate-star-34,1.0,1.0,1.0,1.0
karate-twoStar-34,1.0,0.5,1.0,0.5
karate-cycle-34,1.0,1.0,1.0,1.0
karate-twoClique-34,0.0,0.0,1.0,1.0
karate-complete-34,0.0,0.0,1.0,1.0
\end{filecontents*}
\pgfplotstabletypeset[
      columns={graph,bc_accuracy,cc_accuracy,sv_accuracy,mv_accuracy},
      columns/graph/.style={
          column name=graph $\setminus$ accuracy,
          string type,
      },
      columns/bc_accuracy/.style={
          column name={betweenness},
          postproc cell content/.append code={\pgfkeysalso{@cell content=\pgfmathparse{##1*100}\pgfmathprintnumber{\pgfmathresult}\%}}
      },
      columns/cc_accuracy/.style={
          column name={closeness},
          postproc cell content/.append code={\pgfkeysalso{@cell content=\pgfmathparse{##1*100}\pgfmathprintnumber{\pgfmathresult}\%}}
      },
      columns/sv_accuracy/.style={
          column name={Shapley value},
          postproc cell content/.append code={\pgfkeysalso{@cell content=\pgfmathparse{##1*100}\pgfmathprintnumber{\pgfmathresult}\%}}
      },
      columns/mv_accuracy/.style={
          column name={Myerson value},
          postproc cell content/.append code={\pgfkeysalso{@cell content=\pgfmathparse{##1*100}\pgfmathprintnumber{\pgfmathresult}\%}}
      },
      every last column/.style={column type/.add={}{|}},
  ]{weighted_karate_summary.csv}
\end{table}

\subsection{Experiments to examine correlation of network properties and opinion dynamics}
\label{sec:experiments-description}

We have done simulations of opinion dynamics in one-layer Zachary's karate club network following BVM and two-layer network following GCVM with Zachary's karate club network being an external layer and  different internal network structures. There is a list of internal layers we use in our analysis:
\begin{enumerate}
  \item [1.] {\bf karate}: Zachary's karate club network;
  \item [2.] {\bf star}: star structure with node 0 being the center;
  \item [3.] {\bf two-star}: two central nodes 0 and 17, nodes 1--16 are linked with node 0, nodes 18--33 are linked with node 17. Moreover, nodes 0 and 17 are linked;
  \item [4.] {\bf cycle}: node 0 is linked with node 1, node 1 is linked with node 2, and so on. Finally, node 33 is linked with node 0;
  \item [5.] {\bf two-clique}: nodes 0--16 belong to the first clique, nodes 17--33 --- to the second clique, and these two cliques are connected through link between nodes 0 and 17;
  \item [6.] {\bf complete}: all nodes are linked with each other.
\end{enumerate}

We start with simulations of opinion dynamics of BVM for one-layer network, and then CVM and GCVM for two-layer networks implementing different internal structures and examine their affect on consensus time and winning rate. Network properties (in this work, centrality measures) will also change with changes of the network structure.

\begin{figure}[!htbp]
	\begin{subfigure}{.475\linewidth}
		\includegraphics[width=\linewidth, clip, trim=0cm 0cm 0cm 0.8cm]{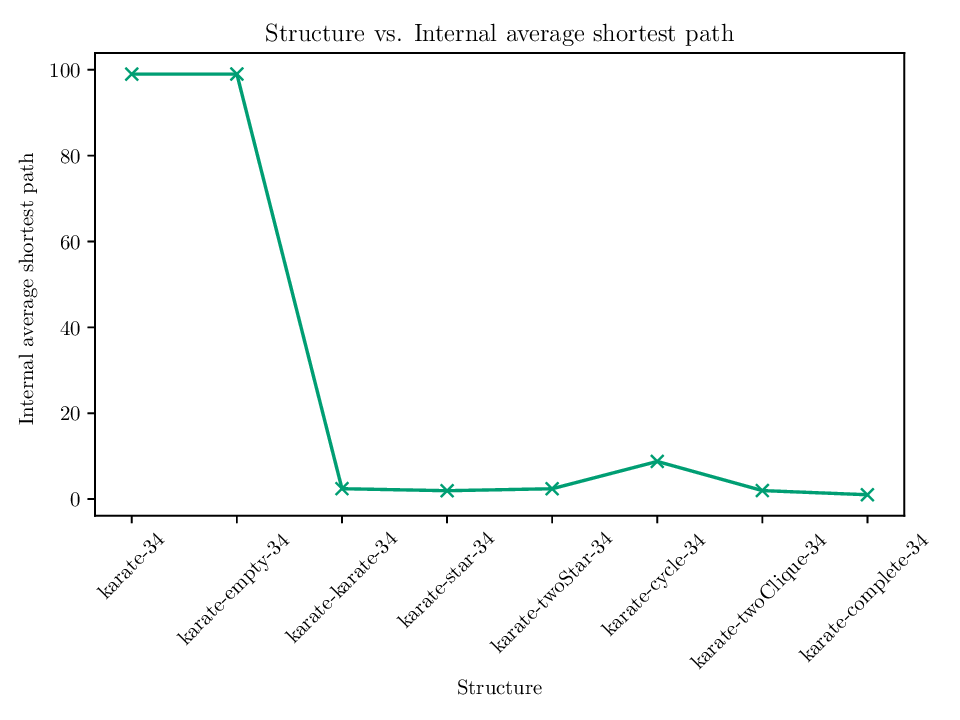}
		\caption{structure vs average shortest path}
		\label{fig:structure-d}
	\end{subfigure}\hfill 
	\begin{subfigure}{.475\linewidth}
		\includegraphics[width=\linewidth, clip, trim=0cm 0cm 0cm 0.8cm]{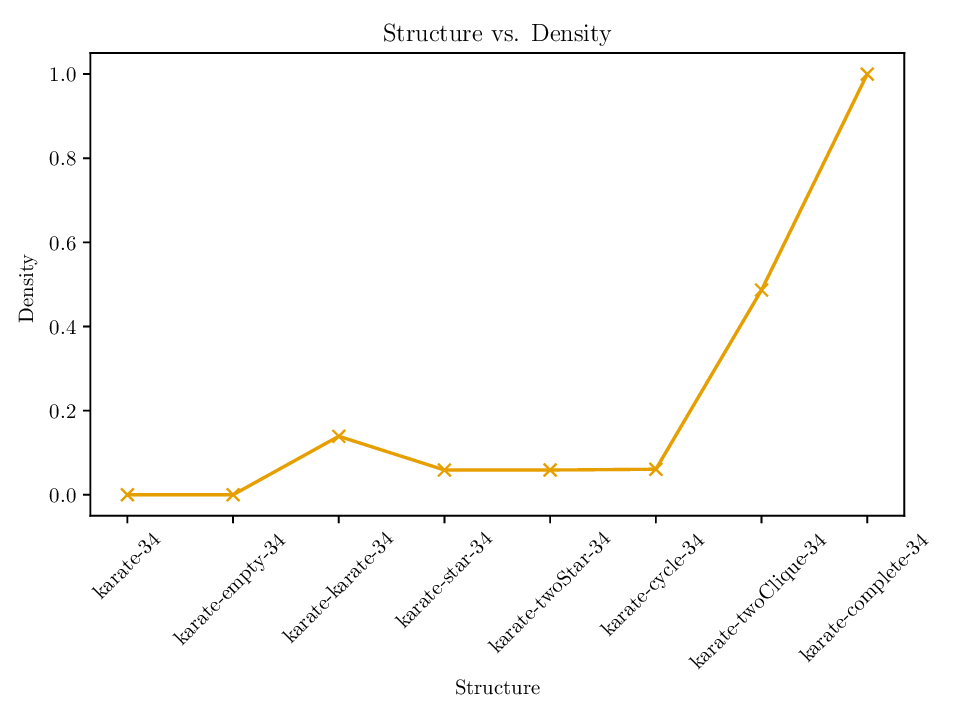}
		\caption{structure vs density}
		\label{fig:structure-s}
	\end{subfigure}\hfill 
	\caption{Internal average shortest path and density for different network structures}
	\label{fig:structure-d-s}
\end{figure}

\begin{figure}[!htbp]
	\includegraphics[width=\textwidth, clip, trim=0cm 0cm 0cm 0.7cm]{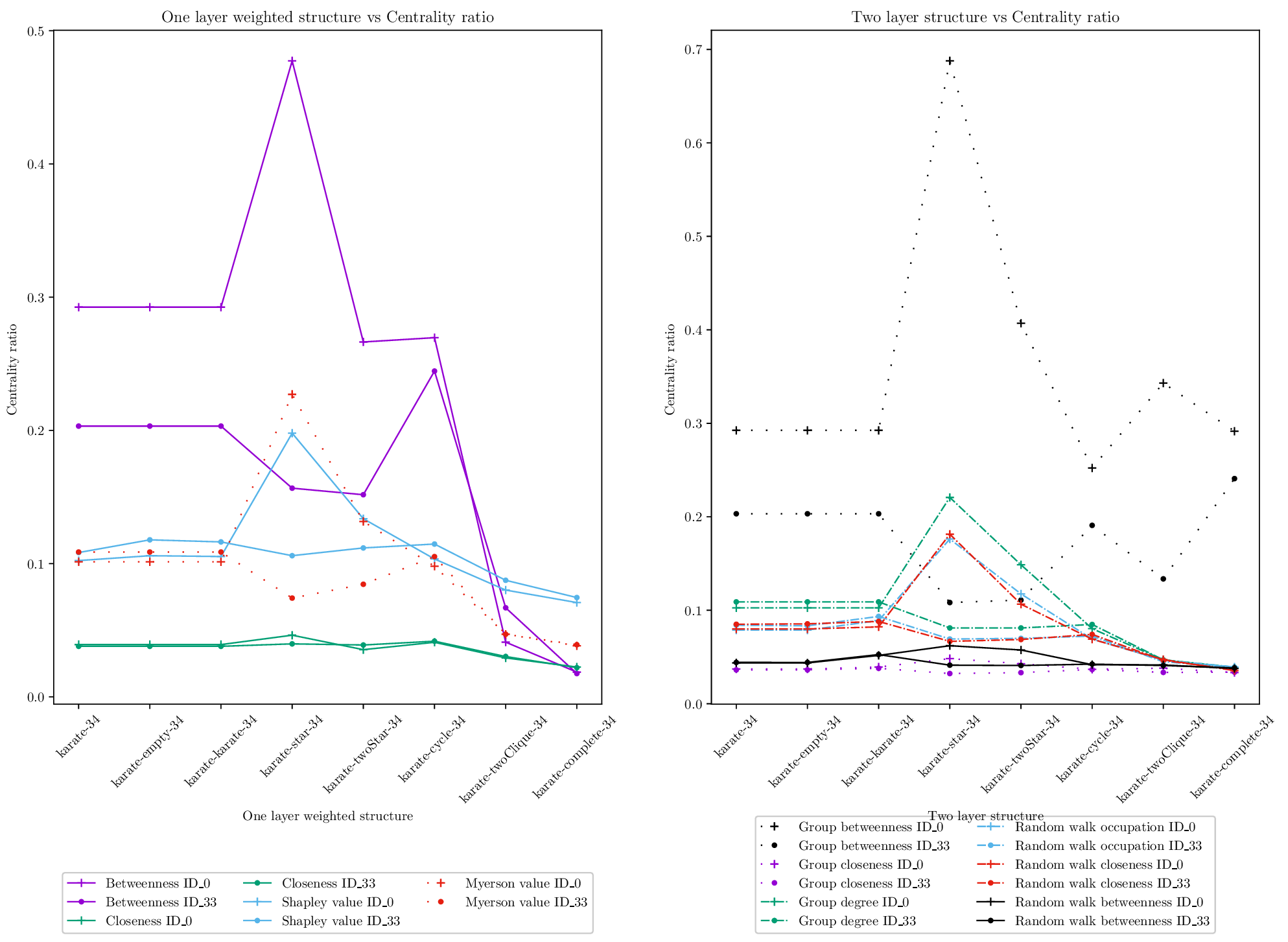}
	\caption{Different centralities for different structures} 
	\label{fig:combination_0_33}
\end{figure}

\begin{figure}[hbt!]
  \begin{subfigure}{.475\linewidth}
    \includegraphics[width=\linewidth, clip, trim=0cm 0cm 0cm 0.8cm]{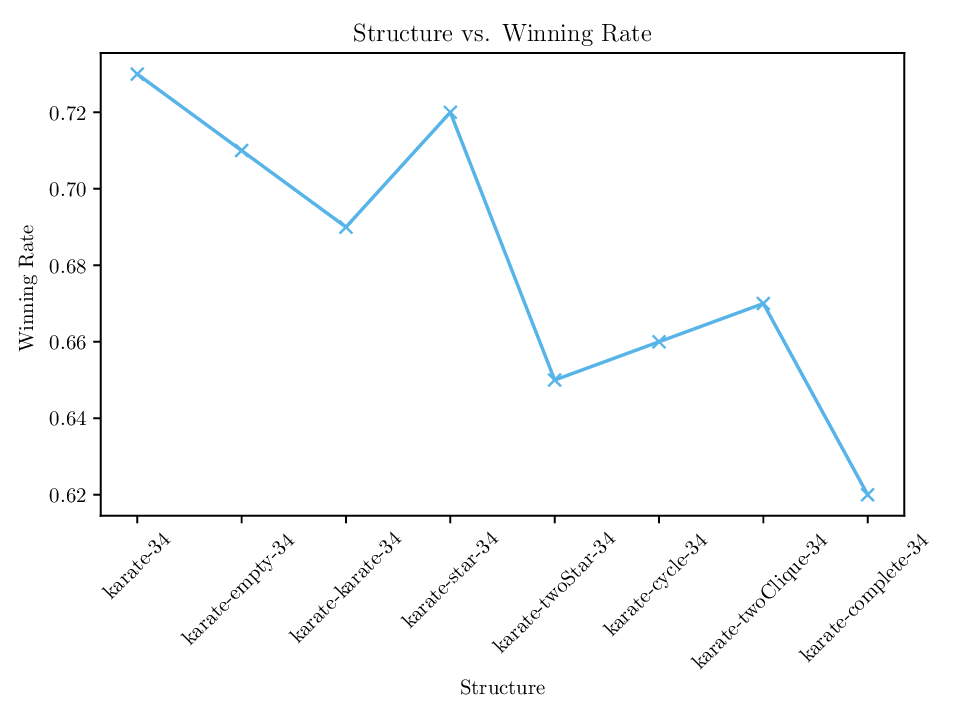}
    \caption{structure vs winning rate}
    \label{fig:structure-wr}
  \end{subfigure}\hfill 
  \begin{subfigure}{.475\linewidth}
    \includegraphics[width=\linewidth, clip, trim=0cm 0cm 0cm 0.8cm]{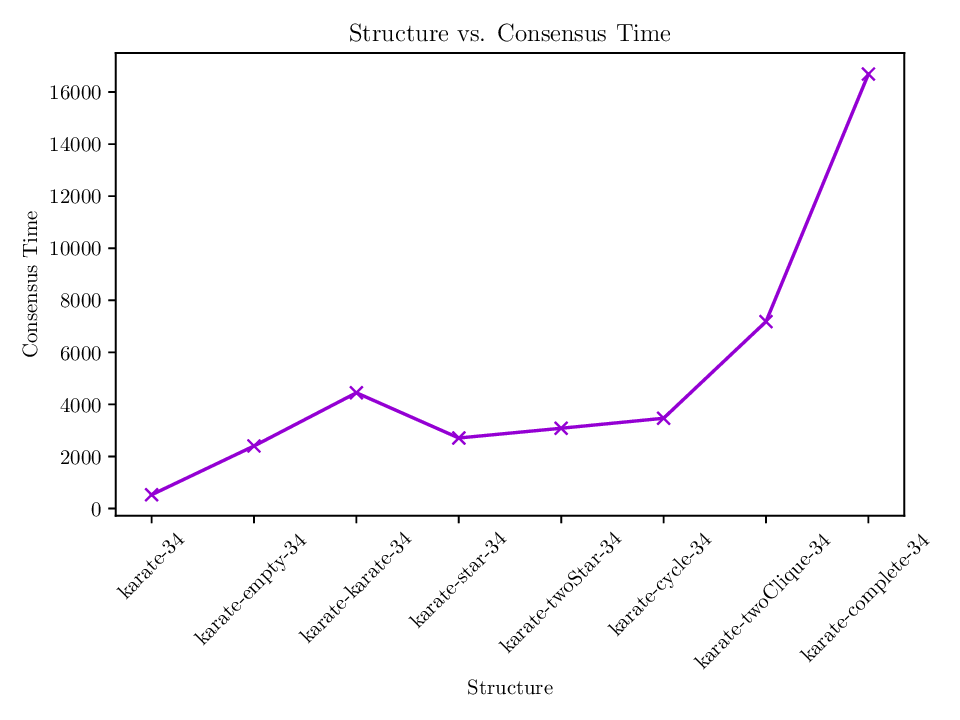}
    \caption{structure vs consensus time}
    \label{fig:structure-ct}
  \end{subfigure}\hfill 
  \caption{Winning rate and consensus time for different structures}
  \label{fig:kpis-structures}
\end{figure}

We describe the results of our experiments:
\begin{itemize}
\item Fig.~\ref{fig:structure-d} shows how internal average shortest path $d_I$ varies depending on the network structure.\footnote{``Karate-34'' and ``karate-empty-34'' refer to a one-layer Zachary's karate club network and to a two-layer network with Zachary's karate club network in external layer and empty internal layer respectively, i.e. $d_I$ does not exist for these two structures, in particular, it is equal to infinity. But in Fig.~\ref{fig:structure-d}, we replace infinity by number $99$.} 
\item Fig.~\ref{fig:structure-s} shows how internal density varies depending on the network structure. By comparing   Fig.~\ref{fig:structure-d} and \ref{fig:structure-s}, we can notice that the internal density $D_I$ and internal average shortest path $d_I$ are negatively correlated.
\item Fig.~\ref{fig:combination_0_33} shows different centralities for Nodes 0 and 33 for different network structures. 
In the left part of Fig.~\ref{fig:combination_0_33}, we present the betweenness centrality, closeness centrality, approximated Shapley value, and approximated Myerson value for Nodes 0 and 33 on the simplified one-layer weighted network with weights calculated by formula \eqref{eq:fusion-rule}. In the right part of Fig.~\ref{fig:combination_0_33}, we present the group degree centrality, group closeness centrality, group betweenness centrality, and two different random walk centralities for the two-layer network with different internal structures. By comparing the centrality trend of Node 0 and Node 33 in the left and right parts of Fig.~\ref{fig:combination_0_33}, respectively, we can observe that both have a similar trend, and this result demonstrates the validity of our approach of simplifying the two-layer network to a one-layer weighted network.
\item Fig.~\ref{fig:kpis-structures} shows how KPIs vary with different network structures. Looking at Fig.~\ref{fig:combination_0_33}, we may notice that some centralities trend of Node 33 are very similar to the trend in Fig.~\ref{fig:structure-d}.  Fig.~\ref{fig:structure-s} also demonstrates the similar trend as in Fig.~\ref{fig:structure-ct}.
\item Fig.~\ref{fig:structure-wr} shows that network structure has a great impact on winning rate.
\item From Fig.~\ref{fig:structure-ct} and \ref{fig:structure-s}, we can notice that there exists a relationship between internal density $D_I$ and consensus time $T_{cons}$. Networks with higher density, like ``karate-complete-34'', take more time to reach consensus, while networks with less density, like ``karate-empty-34'', reach consensus faster. This can be explained by the fact that networks with higher density have more connections which makes it more difficult for a single opinion to dominate quickly.
\end{itemize} 

The main conclusions from the above results are: (i) there is a negative correlation between internal average shortest path $d_I$ and both internal density $D_I$ and consensus time $T_{cons}$, (ii) the approach of simplifying the two-layer network into a one-layer weighted network is valid.

Our next step is to examine dependence of network structure and KPIs of opinion dynamics. To do this, we conducted correlation tests on the above observation results by SciPy \cite{scipysta11:online,2020SciPy-NMeth}. We calculated three coefficients for each pair of characteristics: Pearson \cite{10.1093/biomet/6.2-3.302},  Kendall \cite{kendall1945treatment}  Spearman \cite{spearman1987proof} correlation coefficients. They are presented in Table~\ref{tab:correlation}, where $\ast$, $\ast\ast$, $\ast\ast\ast$ represent a level of significance 0.05, 0.01, 0.001 of the correlation coefficient, respectively.\footnote{We choose node $33$ as the input for centrality. Actually, if we choose another node as an input, the conclusions are still valid.}

We can make several conclusions from Table~\ref{tab:correlation}:
\begin{enumerate}
    \item \textit{High positive significant correlation:} (i) The internal density $D_I$ exhibits a very strong highly significant positive correlation with consensus time across all correlation coefficients (Pearson, Kendall, and Spearman), they all are larger than $0.95$. This indicates that when $D_I$ increases, consensus time $T_{cons}$ significantly increases;  (ii) The internal average shortest path $d_I$ is significantly positively correlated with most centrality measures which is confirmed by Kendall and Spearman correlation coefficients, they are significant.
    
    \item \textit{Negative significant correlation:} $T_{cons}$ shows strong and significant negative correlation with network centrality measures like Betweenness and Closeness, especially with Closeness (with Pearson correlation coefficient equal to $-0.928$). So we can conclude that higher centrality scores are associated with shorter consensus time $T_{cons}$, which is expected. 
    
    \item \textit{Variability in correlations:} Different metrics show varying levels of correlation strength across the Pearson, Kendall, and Spearman correlation coefficients. This variability indicates that the strength and significance of correlations can depend on the correlation method used, likely influenced by the underlying data distributions.
\end{enumerate}

To sum up, Table~\ref{tab:correlation} highlights significant relationships between specific network properties and characteristics of opinion dynamics like consensus time. Centrality of authoritative nodes and network density play a crucial role for consensus time.

\begin{table}[!htbp]
  \centering
  \caption{Correlation coefficients}\label{tab:correlation}
  \begin{tabular}{@{}lccc@{}}
    \toprule
     & Pearson  & Kendall  & Spearman  \\
    \midrule
    $d_I$ vs Betweenness & 0.414 & 0.654* & 0.810* \\
    $d_I$ vs Closeness & 0.231 & 0.192 & 0.270 \\
    $d_I$ vs Shapley value & 0.363 & 0.618* & 0.766* \\
    $d_I$ vs Myerson value & 0.548 & 0.808** & 0.908** \\
    $d_I$ vs Group betweenness & 0.366 & 0.185 & 0.157 \\
    $d_I$ vs Group closeness & 0.408 & 0.333 & 0.614 \\
    $d_I$ vs Group degree & 0.610 & 0.830** & 0.933*** \\
    $d_I$ vs Random walk occupation & 0.475 & 0.691* & 0.826* \\
    $d_I$ vs Random walk closeness & 0.553 & 0.691* & 0.826* \\
    $d_I$ vs Random walk betweenness & 0.161 & 0.618* & 0.778* \\
    \hline
    $T_{cons}$ vs Betweenness & -0.841** & -0.491 & -0.537 \\
    $T_{cons}$ vs Closeness & -0.928*** & -0.340 & -0.464 \\
    $T_{cons}$ vs Shapley value & -0.872** & -0.286 & -0.452 \\
    $T_{cons}$ vs Myerson value & -0.791* & -0.491 & -0.659 \\
    $T_{cons}$ vs Group betweenness & 0.407 & 0.255 & 0.299 \\
    $T_{cons}$ vs Group closeness & -0.291 & 0.109 & 0.036 \\
    $T_{cons}$ vs Group degree & -0.807* & -0.593* & -0.771* \\
    $T_{cons}$ vs Random walk occupation & -0.785* & -0.429 & -0.548 \\
    $T_{cons}$ vs Random walk closeness & -0.834* & -0.357 & -0.524 \\
    $T_{cons}$ vs Random walk betweenness & -0.424 & -0.500 & -0.571 \\
    \hline
    $D_I$ vs $T_{cons}$ & 0.983*** & 0.964** & 0.988*** \\
    \bottomrule
    \end{tabular}
\end{table}

\newpage
\section{Conclusions and future work}

We proposed two fast and accurate algorithms to calculate centrality measures based on game-theoretic approach. Our algorithms can efficiently  approximate the theoretical values of these measure for the networks, for which exact values are computationally difficult to find. Both of our algorithms can identify the most important nodes in the network, which is tested on different examples. The ideas of finding approximated centrality measures in the graphs implemented in our algorithms can be easily transferred to other fields, such as explainable artificial intelligence.

We examined the correlation of several characteristics of opinion dynamics (including BVM, CVM, and GCVM) realized on two-layer networks with characteristics of these networks. As a network in one layer we consider a Zachary's karate club. We examined how internal network structure affects consensus time and winning rate, and if these key performance indicators correlate with network centrality measures.

We find the following developments of our work interesting:
(i) improve the accuracy of the algorithms presented in the paper by improving sampling procedures and to test them on large networks, (ii) exploring the reasons of winning rate variation with the changes in network structure, (iii) obtain analytical expressions (in expectation sense) for approximated Shapley/Myerson values proposed in this paper.

\bibliographystyle{unsrtnat}
\bibliography{mybibliography}  






\end{document}